\documentclass[11pt]{article}
\pdfoutput=1
\usepackage{amsmath}
\usepackage{amssymb}
\usepackage{times}
\usepackage{color}
\usepackage{epsfig,graphics}
\usepackage{bm}
\usepackage{mathrsfs}
\usepackage{enumerate}
\usepackage{authblk}
\usepackage{indentfirst}
\begin{document}

\pagestyle{plain}
\hsize = 6.5 in
\vsize = 8.5 in
\hoffset = -0.75 in
\voffset = -0.5 in
\baselineskip = 0.29 in

\newcommand{\DF}[2]{{\displaystyle\frac{#1}{#2}}}

\title{Unifying deterministic and stochastic ecological dynamics via a landscape-flux approach}

\author[a]{Li Xu}
\author[b,c]{Denis Patterson}
\author[d]{Ann Carla Staver}
\author[e,*]{Simon Asher Levin}
\author[f,*]{Jin Wang}

\affil[a]{State Key Laboratory of Electroanalytical Chemistry, Changchun Institute of Applied Chemistry, Chinese Academy of Sciences, Changchun, Jilin, 130022, P.R. China}
\affil[b]{High Meadows Environmental Institute, Princeton University, Princeton, NJ 08544}
\affil[c]{Department of Mathematics, Brandeis University, Waltham, MA 02454}
\affil[d]{Department of Ecology and Evolutionary Biology, Yale University, New Haven, CT 06520}
\affil[e]{Department of Ecology and Evolutionary Biology, Princeton University, Princeton, NJ 08544}
\affil[f]{Department of Chemistry, Physics and Applied Mathematics, State University of New York at Stony Brook, Stony Brook, NY 11794-3400, USA}
\affil[*]{To whom correspondence should be addressed. E-mail: slevin@princeton.edu, jin.wang.1@stonybrook.edu}

\date{\today}

\maketitle

\begin{abstract}
We develop a landscape-flux framework to investigate observed frequency distributions of vegetation and the stability of these ecological systems under fluctuations. The frequency distributions can characterize the population-potential landscape related to the stability of ecological states. We illustrate the practical utility of this approach by analyzing a forest-savanna model. {\it Savanna}, and {\it Forest} states coexist under certain conditions, consistent with past theoretical work and empirical observations. However, a new {\it Grassland} state, unseen in the corresponding deterministic model, emerges as an alternative quasi-stable state under fluctuations, providing a novel theoretical basis for the appearance of widespread grasslands in some empirical analyses. The ecological dynamics are determined by both the population-potential landscape gradient and the steady-state probability flux. The flux quantifies the net input/output to the ecological system and therefore the degree of nonequilibriumness. Landscape and flux together determine the transitions between stable states characterized by dominant paths and switching rates. The intrinsic potential landscape admits a Lyapunov function, which provides a quantitative measure of global stability. We find that the average flux, entropy production rate, and free energy have significant changes near bifurcations under both finite and zero fluctuation. These may provide both dynamical and thermodynamic origins of the bifurcations. We identified the variances in observed frequency time traces, fluctuations and time irreversibility as kinematic measures for bifurcations. This new framework opens the way to characterize ecological systems globally, to uncover how they change among states, and to quantify the emergence of new quasi-stable states under stochastic fluctuations.
\end{abstract}
\maketitle

\section*{Significance}
Characterizing stability and dynamics of ecological systems under fluctuations is a longstanding challenge in ecology. We study the ecodynamics of a forest-savanna model under fluctuations via a landscape-flux theoretical framework from nonequilibrium statistical physics, and show that ecological dynamics are determined by both population landscape gradients and steady-state probability fluxes. {\it Savanna} and {\it Forest} states coexist under certain conditions, and a {\it Grassland} state unseen in deterministic cases emerges under fluctuations. The intrinsic landscape is identified with a Lyapunov function for quantifying global stability of ecological systems. We quantify barrier heights, kinetic paths, and switching rates between stable states. Average flux, entropy production rate, time irreversibility, variances in time traces, and fluctuations serve as markers to quantify onset/offset of bifurcations.

\section*{Introduction}
The characterization of the dynamics of systems through the construction of sets of differential equations and the exploration of their long-term behavior have become standard and powerful tools in the applied sciences~\cite{VanKampen}. This is particularly the case in mathematical biology and ecology, from the early works of Volterra up to recent years~\cite{Volterra}. The framework of nonlinear dynamical systems allows us to efficiently characterize steady states, periodic orbits, and even more complex invariant sets, as well their dependence on system parameters~\cite{Jorgensen,Wang2015AP,Levin2018PNAS,Staver2011Ecology,Gallopin2006,staver2011global}. However, these methods often rely on local properties of the systems in the vicinity of attractors and are not designed to address how the dynamics are altered in the presence of noise or how fluctuations can induce global switching between multiple stable attractors. In many ecological models, multiple alternative stable states arise, each with their own distinct basin of attraction~\cite{Jorgensen,Gallopin2006}. Characterizing the stability of such states, their responses to ecological fluctuations, and predicting possible transitions between them is a central challenge in ecology. Indeed, on longer time scales, slow variable evolution, perhaps initiated by fluctuations, can alter the topology of the dynamics, leading to critical transitions or to flickering between states or basins of attraction~\cite{Levin2018PNAS,Staver2011Ecology,VanKampen,Gillespie1977JPC,ThomBook1972,Patterson2020SIAM}. We present how landscape-flux theory from non-equilibrium statistical mechanics can provide a powerful framework to study these questions, using a well-known savanna-forest model, typically referred to as the Staver-Levin model (SL model), as a case study~\cite{Levin2018PNAS,Staver2011Ecology}. Here, savanna refers to a grass dominated state with some trees and saplings while forest refers to a tree dominated state with few grass and saplings.

Ecological systems are subject to multifarious sources of stochastic noise, ranging from fires and climatic variability to variations in the growth and death rates of species (see~\cite{Levin2018PNAS,Staver2011Ecology,Scheffer2011Science,Dantas2015EL,staver2011global} and the references therein). In ecological dynamics, the observables of interest fluctuate, which can introduce unpredictability into an otherwise deterministic process. In such settings, the statistics of vegetation or their distributions can be collected and often provide a reliable quantification of the state of an ecological system. The statistical patterns are typically relatively regular and the associated probabilistic dynamics may be predicted since they typically follow the linear evolution law dictated by the associated Fokker-Planck or Master equation. By contrast, the individual stochastic trajectories are the result of nonlinear interactions and are more difficult to reliably predict. These trajectories can be compared with observations of not only the time traces, but also their frequency or statistical distributions, which are increasingly becoming available~\cite{Xu2014PLOSONE}. A critical question is then how observed frequency distributions are linked to stability~\cite{StaverNP2018}. Ecological analyses of snapshot data have mostly assumed that stable states will appear more frequently than unstable ones \cite{Scheffer2003TEE}. However, the real situation can be more complicated, and it remains a challenge to disentangle the underlying stable configurations and stochastic effects.

Historically, researchers have applied a myriad of techniques to study the global dynamics of complex systems subject to random forcing. Some approaches have focused on the zero-noise limit~\cite{Freidlin1984book,Graham1989,Hu1995,Nolting2016Ecology}. Several other approaches considered the finite stochastic fluctuations \cite{VanKampen,Gillespie2000,Ao2004JPA,Zhou2016JCP}
while certain mathematical challenges were posted\cite{Ao2004JPA,Zhou2016JCP}. In this study, we highlight a recently developed method for analyzing the global stability and dynamics of both deterministic and stochastic complex systems in a unified framework~\cite{Wang2015AP,Xu2014PLOSONE,Wang2008PNAS,Wang2011PNAS}. For general dynamical systems, one can study the global dynamics by identifying the so-called population-potential landscape and the rotational curl flux. The global population-potential landscape is determined by the steady-state probability distribution of ecological states, while the rotational curl flux is determined by the steady-state probability flux. Heuristically, the population-potential landscape attracts the system to the basins of the steady states and the curl flux drives the system in a rotational way that reinforces the stability of the flow. Furthermore, since the population-potential landscape is directly
associated with the statistical distribution of the steady states, we can use the observed frequency distribution to infer the underlying population-potential landscape. The observed frequency distributions are typically univariate, allowing an effective description of the ecological system in this specific dimension~\cite{staver2011global}. The corresponding landscape inferred from the observed univariate frequency distribution can provide insights regarding the global stability of the system. If different univariate frequency distributions are observed, one may approximate the multivariate population landscape as the product of the individual univariate distributions (mean field approximation) under the weak coupling assumption. To obtain more precise information on the population landscape, one needs to know the time series of all the observables for the joint distribution, which is rarely the case in practice.

Since the foundational work of Volterra~\cite{Volterra}, a significant amount of research has focused on trying to find Lyapunov functions for dynamic ecological models~\cite{Hastings}, but such approaches have practical difficulties and are still somewhat incomplete~\cite{Wang2015AP,Xu2014PLOSONE,Wang2008PNAS}. The landscape-flux framework provides a general way to analyze the dynamics of ecological systems. We focus on its implications in a forest-savanna model to study the stochastic dynamics and the interplay between grass, saplings, and trees~\cite{Levin2018PNAS,Staver2011Ecology}. This will allow us not only to quantify the underlying population-potential landscape, link this to the frequency distribution of the observables at long times, and gain insights on the global stability under various conditions, but also to characterize the effects of fluctuations on the dynamics.

This study establishes a link between the observed frequency distribution at long times with the population landscape of the ecological system. Moreover, we will identify the rates of transitions between multiple states, and their dependence on noise levels. In particular, contrasting with the view that fluctuations destabilize steady states, we will show that here, fluctuations can sometimes produce a new state not present in deterministic dynamics. Importantly, we show that the non-zero flux characterizing the net input/output to the ecological system and therefore the degree of nonequilibriumness (the distance away from equilibrium), together with the population-potential landscape, determine the noise-induced transitions between the basins of attraction, both in terms of the dominant paths and rates of transition. These noise-induced transitions are irreversible, in the sense that the dominant forward path from A to B is not the same as the dominant backward path from B to A. This new framework thus opens the way to characterize the ecological system globally, to uncover how they switch between states due to the non-zero flux, and to quantify the emergence of new stable states that are not present in the deterministic dynamics.

In order to study the SL model subject to noise, we first compute the non-equilibrium population-potential landscape ($U$) under finite fluctuations. We then take the zero fluctuation limit to obtain the intrinsic potential landscape, denoted by $\phi_0$. As we show presently, the intrinsic potential landscape is a global Lyapunov function for the ecological dynamics and thus the topological structure of the intrinsic potential landscape provides a quantitative measure for the global stability of the ecological system. In addition, we characterize the quantitative relationship between the driving force of the intrinsic potential gradient and that of the probabilistic flux for the ecological dynamics. We also identify the nonequilibrium free energy as a Lyapunov function for quantifying the global stability of the ecological systems at finite fluctuations. The linkage of the non-equilibrium intrinsic free energy with the different phases and the bifurcations (phase transitions) of the ecological system with respect to changes in the parameters.

Complex ecological systems often involve nonlinear interactions that can lead to a variety of behaviors and transitions between various dynamic regimes. These changes in the qualitative behavior of the system can be described by bifurcations where different stable/unstable states can branch out, meet, or emerge spontaneously~\cite{Landau,Yeomans,VanKampen}. Energy, material, or information exchange can also lead to new phases and bifurcations. In nonequilibrium systems, the flux provides the origin of the entropy production, which is a measure of nonequilibrium thermodynamic cost. This cost is a thermodynamic measure of the free energy consumption or dissipation needed for certain biological functions; it is quantified  by the entropy production. For example, in order to maintain the cell cycle flow, nutrition supply through the phosphorylation reaction via ATP hydrolysis, quantified by the entropy production or free energy cost, is required. In an equilibrium ecological system, the bifurcations are determined exclusively by the potential gradient. However, in nonequilibrium ecological systems, the bifurcations are determined by both the potential gradient and the rotational flux. Therefore, the curl flux plays a crucial role in the emergence of nonequilibrium states and bifurcations in nonequilibrium systems~\cite{Xu2020JPCB,Wang2015AP,Fang2019RMP}.

We analyze the average curl flux, {\it Flux$_{av}$}, and entropy production rate,  {\it EPR}, under both finite and zero fluctuations. As system parameters vary, {\it Flux$_{av}$} and the associated thermodynamic cost in terms of the {\it EPR} both have significant changes near (between) the two saddle-node bifurcations, especially in the zero fluctuation case. Therefore, the dynamical and thermodynamic origins of the bifurcation for nonequilibrium ecological systems may be from the curl flux and the {\it EPR} respectively. On the other hand, these physical quantities can be inferred from the observed time series. For example, information on flux and {\it EPR} can be inferred directly from the time irreversibility of the observed time traces. The variance in the frequency statistics and kinetic time obtained directly from the observed time traces can be used as the kinematic markers for the onset or offset of bifurcations. Both physical and kinematic markers based on the observed time traces may be used to identify the start or end of bifurcations in ecological dynamics.
\section*{Materials and methods}
\subsection*{The Staver-Levin Model}
The SL model was introduced to study the dynamics of ecological system and the interplay between the fractions of terrain covered by grass ($G$), savanna saplings ($S$) and savanna trees ($T$) in a forest-savanna ecological system~\cite{Levin2018PNAS,Staver2011Ecology}. In the absence of forest trees, the interaction between savannas and grass is mediated by fires, carried by grass, that limit the rate of maturation of savanna saplings into adult trees. More precisely, the  simplified interactions between grass and two life stages of savanna trees are given by the equations~\cite{Levin2018PNAS,Staver2011Ecology}:
\begin{align}\label{savanna_equation}
\dot{G}&=\mu S+\nu T-\beta GT \nonumber\\
\dot{S}&=\beta GT-(\omega(G)+\mu)S \\
\dot{T}&=\omega(G)S-\nu T.\nonumber
\end{align}
Parameter interpretations and their default values (unless otherwise specified) are given in Table \ref{table1} below. The function $\omega(G)=\omega_0+\frac{\omega_1-\omega_0}{1+e^{-(G-\theta_1)/ss_1}}$
is a smooth decreasing sigmoid accounting for how fires limit the maturation of saplings into adult trees; this aspect of the model can be motivated by percolation theory as well as empirical observations~\cite{schertzer2015implications}.
\begin{table}
\scriptsize
\centering
\caption{Parameter Interpretations and Default Values}
\begin{tabular}{clc}
Symbol & Ecological interpretation & Default  \\
 \hline
$\beta$ & Savanna sapling birth rate & 0.38  \\
$\mu$ & Savanna sapling mortality rate & 0.2  \\
$\nu$ & Adult savanna tree mortality rate & 0.1  \\
$\omega_0$ & Savanna sapling-to-adult recruitment & 0.9 \\
 & rate basic value &   \\
$\omega_1$ & Savanna sapling-to-adult recruitment &  0.2 \\
&  rate of sigmoid basic value &  \\
$\theta_1$ & Grass cover basic value & 0.4  \\
${ss_1}$ & Slope of the sigmoid & 0.01  \\
 \hline
\end{tabular}\label{table1}
\end{table}
\noindent{The SL model assumes that all terrain is covered by one of grass, saplings or savanna trees, so that $G+S+T=1$ for all times. Hence we can reduce the system to a two-dimensional system in which we keep track only of grass ($G$) and savanna trees ($T$) since saplings ($S$) will be given by $S=1-T-G$.}
\subsection*{Population-potential landscape and flux quantification for the SL model}
Due to fluctuations from internal and external sources, the deterministic dynamics described by a set of ordinary differential equations need to be modified to include the contribution of the additional fluctuation forces. Thus the following stochastic dynamics emerge~\cite{Gillespie1977JPC,Wang2015AP,Haken1987,Swain2002PNAS,Fang2019RMP}: $d \mathbf{x} = \mathbf{F}(\mathbf{x})dt + \mathbf{g} \cdot d \mathbf{W} $, where $\mathbf{x}$ is the state vector representing the observables such as population or species density, $\mathbf{F}(\mathbf{x})$ is the driving force for the dynamics, and $\mathbf{W}$ coupled with the matrix $\mathbf{g}$ represents an independent Gaussian fluctuating process. We set $D \mathbf{G}=(1/2) (\mathbf{g} \cdot \mathbf{g}^{\mathbf{T}})$, where $D$ is a constant describing the scale of the fluctuations and $\mathbf{G}$ represents the diffusion matrix of the fluctuations. In this study, $\mathbf{G}$ is an isotropic diagonal identity matrix, for simplicity, and thus the noise is chosen as Gaussian white noise.

The stochastic dynamics are characterized by the probability distribution of the system state at time $t$, $P(\mathbf{x},t)$, which can be obtained by solving the Fokker-Planck equation:
\begin{equation}\label{FPE}
\partial_t{P} = -\nabla\cdot\mathbf{J} = -\nabla\cdot[\mathbf{F}P - (1/2) \nabla \cdot ( (\mathbf{g}\cdot\mathbf{g}^{\mathbf{T}}) P )]
\end{equation}
Hence, the steady-state probability distribution, denoted by $P_{ss}$, can be obtained by solving the steady-state Fokker-Planck equation, i.e. ${\partial_t{P}}=0$. In the equilibrium systems, the probability follows a Boltzmann distribution $P \sim \exp[-U]$~\cite{Gillespie1977JPC,VanKampen,Wang2015AP,Prigogine} and the energy $U$ is called the population-potential landscape. Thus the driving force for the dynamics is determined by the gradient of the population-potential landscape in equilibrium systems.

In nonequilibrium systems, the force $\mathbf{F}$ cannot be written as the gradient of a potential in general. However, the population-potential landscape can still be defined and linked to the probabilities by the formula: $U= -\ln P_{ss}$~\cite{Wang2008PNAS,Wang2015AP,VanKampen}. We denote by $\mathbf{J_{ss}} = \mathbf{F}P_{ss} - D\nabla\cdot(\mathbf{G}P_{ss})$ the corresponding steady-state probability flux and note that it satisfies the divergence-free condition $\nabla \cdot \mathbf{J}_{ss}=0$.

In equilibrium systems, there is no net flux in or out of the system. Thus, the steady state probability flux is zero at all points in the state space; this is the so-called detailed balance condition $\mathbf{J}_{ss} = 0$. In non-equilibrium systems, the non-zero flux, $\mathbf{J}_{ss}$, is divergence-free and breaks the detailed balance condition. This non-zero flux thus provides a quantitative measure of the degree to which the system is out of equilibrium. In non-equilibrium ecological systems, the driving force $\mathbf{F}$ can be decomposed into the gradient of the potential $U$, the curl steady state probability flux and the divergence of the diffusion coefficient as $\mathbf{F} = -D\mathbf{G}\cdot\nabla U + \mathbf{J}_{ss}/P_{ss} + D\nabla\cdot\mathbf{G}$. The population potential landscape $U$ and the steady state probability flux $\mathbf{J}_{ss}$ together can address many global dynamical and thermal dynamical issues including stability, robustness, dynamics and thermodynamics of ecological systems. We use the SL model under fluctuations to study the stochastic dynamics and the interplay among grass, saplings and trees~\cite{Levin2018PNAS,Staver2011Ecology}. We solve the Fokker-Planck PDE given by \eqref{FPE} for the SL model with reflecting boundary conditions, i.e. $\mathbf{n}\cdot \mathbf{J}=0$ where $\mathbf{n}$ is a unit vector perpendicular to the boundary of the state-space, to obtain the probability distribution of the system. We can thus quantify the population-potential landscape $U$ and the flux $\mathbf{J}_{ss}$, which together determine the driving forces for the dynamics of the ecological system.

Forest-savanna landscapes are non-equilibrium open ecological systems and hence exchange energy with their environments, which leads to dissipation. The entropy of a stochastic system can be defined as $S_{entropy}= - \int P \ln P d\mathbf{x}$ and the change in the entropy in time can be divided into the entropy
production rate and heat dissipation rate. The time evolution of the entropy of the system is thus given by: $\dot S_{entropy} = \dot S_t - \dot S_e$, where the entropy production rate ({\it EPR}$=\dot S_t$) is given as $\dot
S_t = \int d\mathbf{x} (\mathbf{J} \cdot (D\mathbf{G})^{-1} \cdot \mathbf{J}) /
P$~\cite{Qian2006,Wang2008PNAS,Zhang2012JCP,Ge}. Thus, the {\it EPR} is explicitly linked to the flux $\mathbf{J}$. Zero flux would give rise to zero entropy production, which would correspond to an equilibrium system. However, in practice, non-zero fluxes are likely, corresponding to non-equilibrium systems. A higher flux gives rise to a higher {\it EPR}, corresponding to more deviations from equilibrium. This formalism provides a link between non-equilibrium driving force (flux) and non-equilibrium thermodynamic cost, i.e. the {\it EPR}.

One can prove that $\dot S_t \geq 0$, which leads to the second law of nonequilibrium thermodynamics. $ \dot S_t$ has the physical meaning of the {\it EPR} contributed from both the system $\dot S$ and environment $\dot S_e$ as $\dot S_t = \dot S_{entropy} +  \dot S_e$. This can be understood as a formulation of the first law of nonequilibrium thermodynamics.  The heat dissipation rate from the environment is given as $\dot S_e =  \int d\mathbf{x} (\mathbf{J} \cdot (D\mathbf{G})^{-1} \cdot (\mathbf{F}-D\nabla\cdot\mathbf{G})) $. Heat dissipation rate can be either positive or negative and can quantify the entropy flow rate from the environment to the non-equilibrium system. When the system is at the steady-state $\dot S_{entropy} = 0$, the {\it EPR} and the heat dissipation rate are equal~\cite{Ge,Qian2006, Wang2008PNAS, Zhang2012JCP}. The {\it EPR} and the average flux {\it Flux$_{av}=\int \mathbf{J}d\mathbf{x}$} thus provide global thermodynamic measures for the non-equilibrium systems~\cite{Wang2015AP,Xu2014PLOSONE,Wang2008PNAS,Zhang2012JCP}.
\begin{table*}
\scriptsize
\centering
\caption{Mathematical Variables}
\begin{tabular}{ll}
Symbol & Interpretation \\
\hline
${\bf x}$  &  system state \\
${\bf F}({\bf x})$ &    driving force\\
$D$  &  scale factor representing the magnitude of the fluctuations\\
${\bf G}$  &  diffusion matrix\\
$P ({\bf x},t)$  &  probability of system state ${\bf x}$ at time $t$\\
$ {\bf J}({\bf x},t)$  &  probability flux\\
$J_{ss}$  &  probability flux of steady state\\
$U(\mathbf{x})$  &  population potential landscape\\
$\phi_0$  &  intrinsic potential landscape\\
$\mathbf{V}$  &  intrinsic flux velocity\\
$L({\bf x})$  &  Lagrangian\\
$\mathcal{F}$  &  intrinsic free energy\\
$\mathcal{Z}$  &  partition function\\
\hline
\end{tabular}
\end{table*}
\subsection*{Lyapunov function for the SL model under zero fluctuations}
A Lyapunov function is crucial for quantifying the global stability
of ecological systems subject to perturbations. One might use
the steady state probability or the associated population potential to explore
the global stability under finite fluctuations. However,
the population potential is not a Lyapunov function in
general\cite{Xu2014PLOSONE} and it is often a challenging problem to find Lyapunov
functions for complex non-equilibrium systems. Here we show that the intrinsic potential landscape, $\phi_0$, is a Lyapunov function for the ecological dynamics in the zero noise limit~\cite{Zhang2012JCP,Xu2014PLOSONE}.

The probability $P$ can be expanded according to the
fluctuation strength $D$ as below:
\begin{eqnarray}\label{P_taylor}
P(\mathbf{x})=\exp (-(\phi_0(\mathbf{x})/D+\phi_1(\mathbf{x})+D\phi_2(\mathbf{x})+\cdots))/Z
\end{eqnarray}
where $Z=\int \;\exp (-U(\mathbf{x}))d\mathbf{x}$. By substituting \eqref{P_taylor} into \eqref{FPE}, we obtain the $D^{-1}$ order expansion of the Fokker-Planck equation, which has the largest contribution to the probability under zero noise limit. This yields the Hamilton-Jacobi equation (HJE):
\begin{eqnarray}
H=\mathbf{F} \cdot \nabla \phi_0
+ \nabla \phi_0 \cdot \mathbf{G} \cdot \nabla \phi_0 = 0.
\end{eqnarray}
The time evolution of $\phi_0(\mathbf{x})$ is thus given by:
$\dot \phi_0(\mathbf{x})= \dot {\mathbf{x}} \cdot\nabla
\phi_0
 = \mathbf{F} \cdot \nabla \phi_0
= - \nabla \phi_0 \cdot \mathbf{G} \cdot \nabla \phi_0 \leq 0.$
The value of $\phi_0(\mathbf{x})$ monotonically decreases along the deterministic trajectories under the zero fluctuation limit since $\mathbf{G}$ is positive definite. Therefore, $\phi_0$ is a Lyapunov function and can be used to quantify the global stability of the systems. Furthermore, $\phi_0$ is linked with the steady state probability and population-potential landscape as $U=-lnP_{ss} \sim \phi_0/D$. The solution $\phi_0$ of the Hamilton-Jacobi equation, which is the zero-fluctuation limit of the solution to the Fokker-Plank equation, is called the intrinsic potential of the system~\cite{Zhang2012JCP,Xu2014PLOSONE}.

In the zero fluctuation limit, the force ${\bf F}$ can be decomposed into a gradient term and a curl term: $\mathbf{F}
= -\mathbf{G}\cdot\nabla\phi_0 + (\mathbf{J}_{ss}/P_{ss})|_{D\rightarrow0}=-\mathbf{G}\cdot\nabla\phi_0 + \mathbf{V}$, where $-\mathbf{G}\cdot\nabla\phi_0$ represents the gradient of the non-equilibrium intrinsic potential. We set $\mathbf{V} = (\mathbf{J}_{ss}/P_{ss})_{D \rightarrow 0}$ as the intrinsic steady-state flux velocity. $\mathbf{J}_{ss}|_{D\rightarrow0}$ represents the steady-state intrinsic divergence free curl flux due to $\nabla \cdot \mathbf{V}=0$. From the Hamilton-Jacobi equation, the relationship between $\phi_0$ and the intrinsic flux is thus $(\mathbf{J}_{ss}/P_{ss})|_{D\rightarrow0} \cdot \nabla \phi_0 = \mathbf{V} \cdot \nabla \phi_0= 0$. This implies that the gradient of the intrinsic potential and the intrinsic flux are perpendicular to each other in the zero fluctuation limit.

Due to the normalization condition in the SL model, i.e. $G+S+T=1$, the state space becomes an isosceles triangle, making calculation of $\phi_0$ very difficult. To overcome this problem, we approximate the Lyapunov function $\phi_0$ from the expansion of the potential $U(\mathbf{x})$ for the small diffusion coefficient $D$ as $U(\mathbf{x})=\phi_0(\mathbf{x})/D+\phi_1(\mathbf{x})+...$. We applied the linear fit method for the diffusion coefficient $D$ versus the $DU$ to solve the $\phi_0$ approximately. We use the data of $0.0002<D<0.0005$ to fit a line, which is the diffusion coefficient $D$ versus $D\ln P_{ss}$). Thus, the slope of the line leads to the value of $\phi_0$~\cite{Zhang2012JCP,Yan2013PNAS}. The results are shown later in this paper. An exact numerical solution of the Hamilton-Jacobian equation for the intrinsic potential landscape $\phi_0$  under a specific choice of diffusion matrix mimicking the population evolution dynamics is demonstrated in the SI.
\subsection*{Non-equilibrium thermodynamics, entropy, energy and free energy of the general dynamical systems under the zero-fluctuation limit and the finite fluctuations}
In equilibrium systems, we can quantify the equilibrium probability distribution, the partition function as well as the entropy and free energy according to the underlying interacting potential energy. The partition function provides a statistical description for the collection of states in the system. For non-equilibrium systems, the intrinsic potential $\phi_0$ can be related to the steady-state probability distribution under the zero-fluctuation limit: $\mathcal{P}_{ss}(\mathbf{x}) = P_{ss}(\mathbf{x})|_{D\rightarrow0} = \exp(-\phi_0/\mathcal{D})/\mathcal{Z}$, where $\mathcal{D} = D|_{D\rightarrow0}$. The partition function $\mathcal{Z}$ is defined as $\mathcal{Z} = \int \exp(-\phi_0/\mathcal{D})d\mathbf{x}$. Thus, $\phi_0 = -\mathcal{D} \ln(\mathcal{Z} \mathcal{P}_{ss})$.

The entropy of the non-equilibrium system under the zero-fluctuation limit can be defined by:
$\mathcal{S} = - \int \mathcal{P}(\mathbf{x},t) \ln\mathcal{P}(\mathbf{x},t) d\mathbf{x}$~\cite{Wang2015AP,Zhang2012JCP,Xu2014PLOSONE,Wang2008PNAS,Zhang2012JCP}. The intrinsic energy $\mathcal{E}$ of the non-equilibrium system can be defined as: $\mathcal{E} = \int \phi_0 \mathcal{P}(\mathbf{x},t) d\mathbf{x} = - \mathcal{D} \int \ln (\mathcal{Z} \mathcal{P}_{ss}) \mathcal{P}(\mathbf{x},t) d\mathbf{x}$. Thus, the intrinsic free energy of the non-equilibrium system can be defined as $\mathcal{F} = \mathcal{E} - \mathcal{D} \mathcal{S} = \mathcal{D} \left(\int \mathcal{P} \ln(\mathcal{P}/\mathcal{P}_{ss}) d\mathbf{x} - \ln\mathcal{Z}\right)$.

The non-equilibrium intrinsic free energy always decreases since~\cite{Zhang2012JCP,Xu2014PLOSONE}
\begin{eqnarray}
\frac{d \mathcal{F}}{d t}
= - \mathcal{D}^2 \left( \int \left[ \nabla \ln(\frac{\mathcal{P}}{\mathcal{P}_{ss}}) \cdot \mathbf{G} \cdot \nabla \ln(\frac{\mathcal{P}}{\mathcal{P}_{ss}}) \right] \mathcal{P} d\mathbf{x} \right) \leq 0
\end{eqnarray}.
The minimum value of the non-equilibrium intrinsic free energy is $\mathcal{F} = -\mathcal{D} \ln\mathcal{Z}$. This represents the second law of thermodynamics for non-equilibrium systems. Therefore, the non-equilibrium intrinsic free energy is a Lyapunov function and, as we show presently, it can be used to quantify the global stability of the non-equilibrium system.

We also explore the non-equilibrium free energy under finite fluctuations $D$.
The energy $\mathcal{E}$ of the non-equilibrium system under finite fluctuations can be defined as: $\mathcal{E}=\int D U P d\mathbf{x}$, and the entropy under finite fluctuations is $S_{entropy}= - \int P \mathrm{ln}P d\mathbf{x}$. Thus, the free energy under finite fluctuations is shown as: $\mathcal{F} = \mathcal{E} - D S_{entropy}= D \int P \mathrm{ln}(P/P_{ss}({\bf x})) d\mathbf{x}$\cite{Zhang2012JCP,Xu2014PLOSONE}. The non-equilibrium free energy under finite fluctuations also always decreases since $\frac{d \mathcal{F}}{d t}
= - D^2 \left( \int \left[ \nabla \ln(\frac{P}{P_{ss}}) \cdot \mathbf{G} \cdot \nabla \ln(\frac{P}{P_{ss}}) \right] P d\mathbf{x} \right) \leq 0 $~\cite{Zhang2012JCP,Xu2014PLOSONE}. This shows that free energy under finite fluctuations is also a Lyapunov function\cite{Zhang2012JCP,Xu2014PLOSONE}, which can be used to quantify the global stability of the non-equilibrium system.
\subsection*{Kinetic speed and dominant paths between the stable states}
The path integral approach can be used to identify and quantify the most likely transitions between two stable states. The path probability starts from the initial state
$\mathbf{x}_i$ at $t = 0$, and end at the final state of
$\mathbf{x}_f$ at time $t$. The path integral formula characterizing the probability from the initial state to the final state through the paths is shown as:
$P({\mathbf{x}_{f}},t |{\mathbf{x}_{i}},0)= \int {D} {\bf x} \; \exp[-\int dt (\frac{1}{2} \nabla \cdot {\bf F}({\bf x}) + \frac{1}{4} (d {\bf x}/dt
- {\bf F}({\bf x})) \cdot \frac{1}{{\bf D}({\bf x})} \cdot (d {\bf x}/dt - {\bf F}({\bf x})))]
=\int D {\mathbf{x} } \exp[-A(\mathbf{x})]=\int D {\mathbf{x} } \exp[-\int L({\mathbf{x}} (t))dt]$, where $L({\mathbf{x}} (t))$ is the Lagrangian and $A(\mathbf{x})$ is the action for each path on the potential landscapes. The path integral over $D {\mathbf{x}}$ represents the sum over all possible paths connecting ${\mathbf{x}_{i}}$ at time zero to ${\mathbf{x}_{f}}$ at time $t$. The exponential factor gives the weight of each specific trajectory and the probability of going from ${\mathbf{x}_{i}}$ to ${\mathbf{x}_{f}}$ is thus the weighted sum over all possible paths. The path integral can be approximated by the path that contributes the most to the weight since the other paths contributions are exponentially small. We can find the dominant paths with the optimal weights through minimization of the action $A(\mathbf{x})$ or Lagrangian $L({\mathbf{x}} (t))$ as the dominant path probability is proportional to $\exp[-A(\mathbf{x})]$. Thus, we can identify the paths that give the largest contribution to the weight as the dominant {\it Savanna}-{\it Forest} switching pathways.
\subsection*{Ecological Dynamics: The landscape-flux approach versus conventional nonlinear dynamics}
\newcommand{\tabincell}[2]{\begin{tabular}{@{}#1@{}}#2\end{tabular}}
\begin{table*}
\tiny
\centering
\caption{Comparison between conventional nonlinear dynamics analysis and landscape-flux theory}
\begin{tabular}{l|l}
Conventional  nonlinear dynamics & Landscape-flux approach \\
\hline
locations of the fixed point states  &  weights of the states  \\
local dynamics around fixed points &    global connections between the fixed point states \\
local stability analysis  &  global stability analysis based on landscape and flux under fluctuations \\
local stability analysis  &   Lyapunov function identifications for quantifying global stability under no fluctuations \\
locations of the fixed points   & barrier height between the local stable states \\
locations of the fixed points   & kinetic rates of switching between the local stable states \\
locations of the fixed points   & dominant kinetic pathways and associated weights between local stable states \\
phase diagram and locations of bifurcations &  physical origins and early warning signals of bifurcations \\
phase diagram and locations of bifurcations &  the kinematic markers and early warning signals of the bifurcations \\
no identification of the driving force components  &  theoretical framework of the driving force via landscape and flux \\
  &  new stable state can emerge from the fluctuations \\
\hline
\end{tabular}
\end{table*}
In standard deterministic nonlinear dynamics, local stability analysis can be performed and stable states can be identified. However, there is no information about the weights of the states, information that can be provided from the probabilistic landscape-flux approach. Furthermore, local stability analysis does not quantify the connections or switching paths between the stable states, properties that can be quantified through the landscape-flux approach. In addition, conventional nonlinear dynamical analysis often does not provide information about the global stability of a system, typically due to the difficulty in finding a suitable Lyapunov function. The landscape-flux approach can provide a way to identify the Lyapunov function and therefore quantify the global stability.

The additional information that this approach provides is the degree of difficulty in switching from one stable state to another, a property that can have significant ecological consequences in real-world systems. Similarly, while conventional nonlinear analysis can identify the bifurcations for the system, there is typically no information about how the bifurcations occur and the possible origins of such bifurcations. This is one of the most important unresolved issues in theoretical ecology and can be addressed in the landscape-flux approach by searching for the physical (dynamic and thermodynamic) origins of the bifurcations of the ecological systems.

In deterministic nonlinear dynamics, the driving force in the model is simply the right hand side of the evolution equations. However, there is usually no easy way to understand the nature of this driving force when the system is subject to noise. The landscape-flux approach offers a framework to study the stochastic dynamics by identifying the driving force as the action of both the landscape gradient and rotational flux. One can also quantify the associated global thermodynamics in terms of the {\it EPR}. Therefore, the landscape-flux approach provides a general framework to study the global dynamics and thermodynamics of the ecological systems. Finally, in deterministic nonlinear dynamics, the stable states emerge from the interactions. However, in the presence of stochastic forcing, new quasi-stable states can emerge, and these new states can be predicted by the landscape-flux approach.
\section*{Results}
\subsection*{1.Dynamics and Thermodynamics via Potential-flux Landscapes}
\subsubsection*{1.1.a. Non-equilibrium population-potential landscape and flux with finite fluctuation}
We now illustrate the landscape-flux approach to ecological dynamics by studying the SL model under fluctuations. {\it Savanna} denotes a grass dominated
state with some trees and saplings, while {\it Forest} is a tree dominated state with few grass and saplings. The {\it Grassland} state has no saplings or trees present.

\begin{figure}[!ht]
\centering
\includegraphics[width=0.8\textwidth]{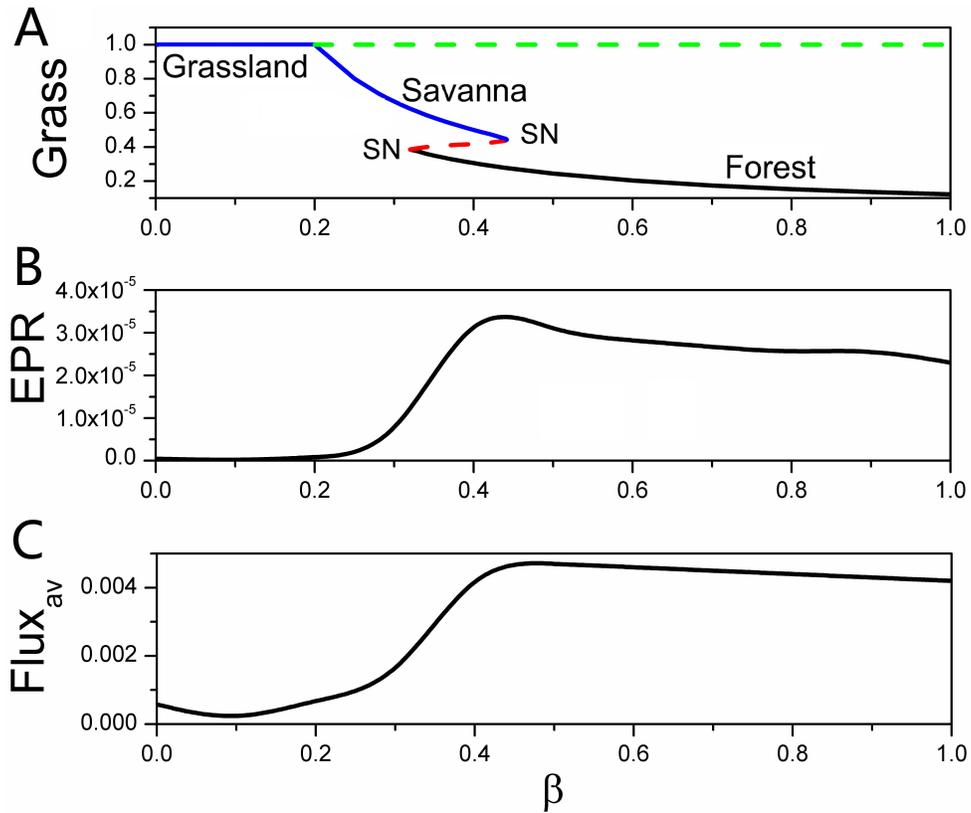}
\caption{ A: The phase diagram  versus $\beta$. B: The population entropy production rate versus $\beta$. C: The population average flux versus $\beta$.}\label{beta_phase}%
\end{figure}

\begin{figure}[!ht]
\centering
\includegraphics[width=0.8\textwidth]{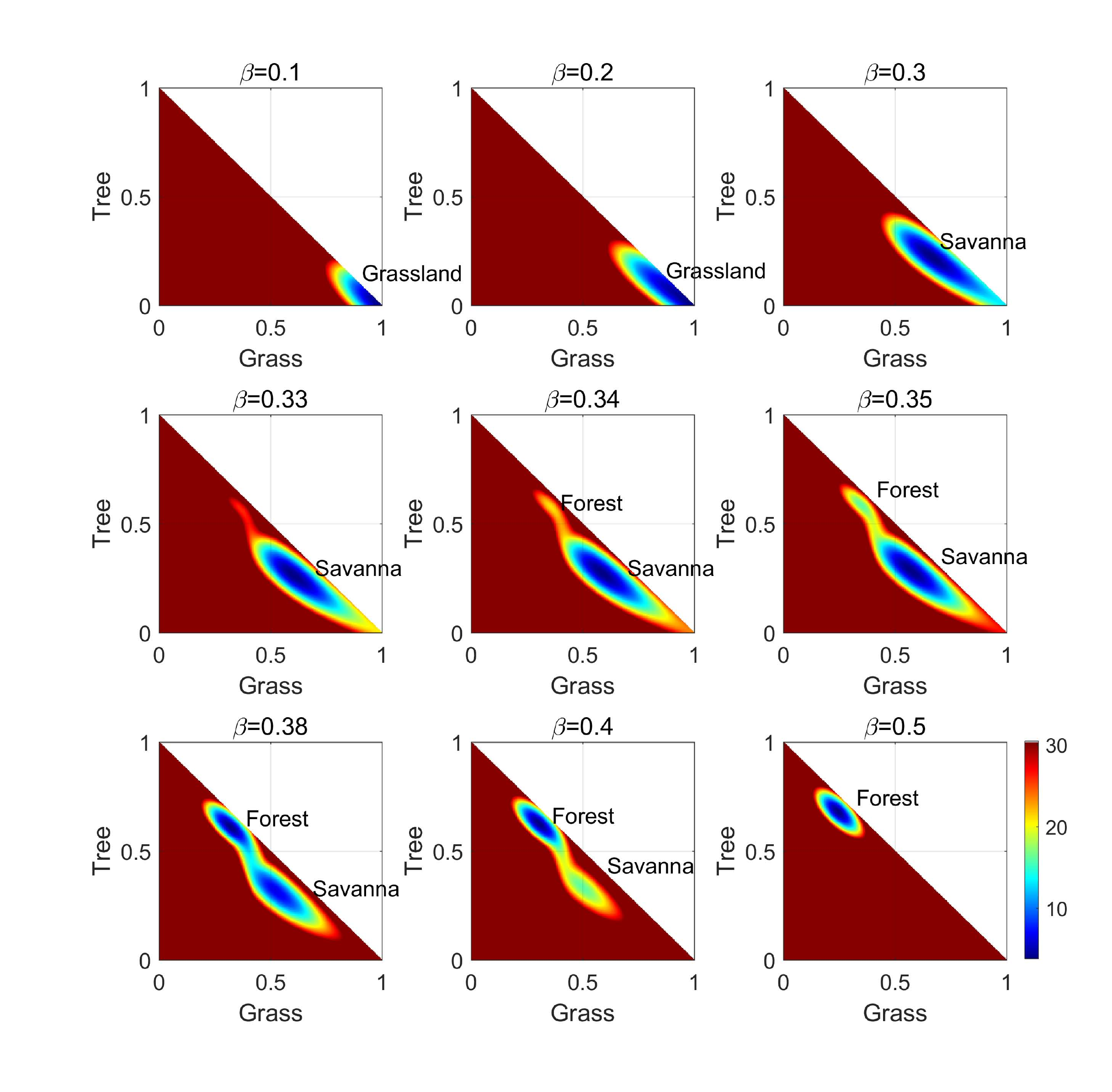}
\caption{ The two-dimensional population-potential landscapes versus $\beta$ with finite fluctuation $D=0.0001$.}\label{betaU2d}%
\end{figure}

\begin{figure}[!ht]
\centering
\includegraphics[width=0.8\textwidth]{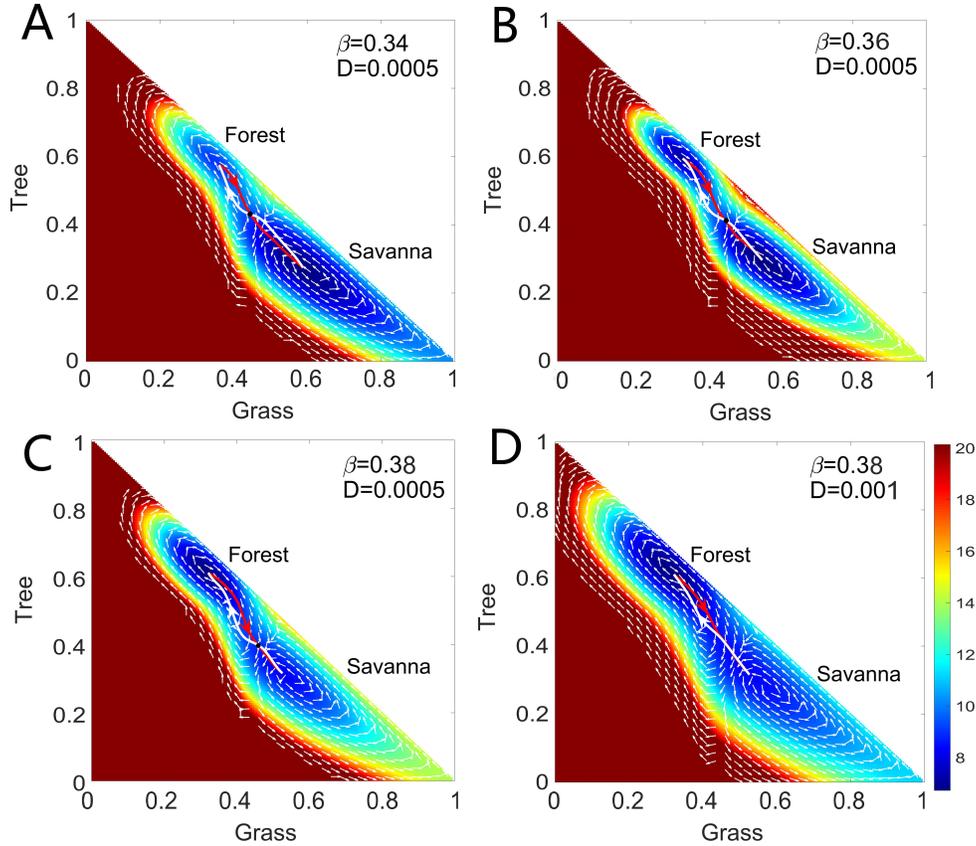}
\caption{A,B,C: The dominant population paths and fluxes on the population-potential landscape $U$ with different $\beta=0.34,0.36,0.38$ and $D=0.0005$. The white lines represent the dominant population paths from the {\it Savanna} state to {\it Forest} state. The red lines represent the dominant population paths from the {\it Forest} state to {\it Savanna} state. The white arrows represent the steady-state probability fluxes. D. The dominant population paths and fluxes on the population-potential landscape $U$ with $\beta=0.38$ and $D=0.001$}\label{betaflux}%
\end{figure}

Figure \ref{beta_phase}A shows the deterministic phase diagram in $\beta$ (the sapling birth rate), while Figure \ref{betaU2d} shows two-dimensional population-potential landscapes ($U$) for a range of $\beta$'s with $D=0.0001$; in both cases the system is considered under finite fluctuations. In Figure \ref{betaU2d}, the population-potential landscape initially has one stable state that evolves from the {\it Grassland} state (i.e. $(G,T) = (1,0)$) to the {\it Savanna} state with increasing $\beta$. As $\beta$ increases further, the stable {\it Forest} state emerges. As $\beta$ increases, the ecological system switches from {\it Savanna} dominant to {\it Forest} dominant (also shown in Figure \ref{beta_phase}A), and as $\beta$ increases further, the grass is completely invaded by the trees due to the high sapling birth rate. Eventually, the {\it Forest} state becomes dominant while the {\it Savanna} state disappears. Figure \ref{betaflux} shows the fluxes on the population-potential landscapes as white arrows for increasing $\beta$. In order to show the fluxes clearly, we only give the directions of the larger values of the fluxes, which are all around the stable states. Vegetative growth factors, such as nutrition and energy from soil, air, water and sunshine, will vary due to different climates in the environment. Thus, when the system has two stable states {\it Savanna} and {\it Forest}, the fluxes originating from the vegetative growth factors go around the stable states, enhancing their communications to each other. In Figure \ref{betaflux}, we can see that the negative gradient of the population-potential landscapes and the non-zero flux are the driving forces of the forest-savanna ecological system.

Figure \ref{betaflux} shows the dominant population paths, calculated under finite fluctuations, on the population-potential landscape $U$ with different parameters: (A) $\beta=0.34,D=0.0005$, (B) $\beta=0.36,D=0.0005$, (C) $\beta=0.38,D=0.0005$, (D) $\beta=0.38,D=0.001$. Once more, {\it Savanna} and {\it Forest} are the two stable states. The red line is the dominant population path from {\it Forest} state to {\it Savanna}, while the thick white line is the dominant population path from {\it Savanna} state to {\it Forest}. The white arrows represent the steady-state probability fluxes which guide the dominant population paths deviating from the naively expected steepest descent path passing through the saddle point based purely on the population-potential landscape. Therefore, the dominant population paths from {\it Savanna} to {\it Forest} and the dominant population paths from {\it Forest} to {\it Savanna} do not follow the same path, in contrast to the equilibrium case under zero flux. The two dominant population paths are different and apart from each other, which shows the irreversibility of the dominant population paths due to the presence of the nonequilibrium rotational flux. This shows that the dominant population paths going from {\it Savanna} to {\it Forest} (the white lines) and going from {\it Forest} to {\it Savanna} (the red lines) will follow different routes. The two dominant population paths under larger fluctuations (bigger diffusion coefficient $D$) are shown in Figure \ref{betaflux}D, the two dominant population paths are also apart from each other due to presence of the steady-state probability fluxes such as those shown in Figure \ref{betaflux}C. The fluxes have spiral shapes around these two basins which show the dynamic nature of non-equilibrium system.

The red lines and the white lines approximately form a figure of ''8'' shape, emphasizing that the switching dominant population paths are distinct and thus that the transition is ''irreversible''. The white path lines start from {\it Savanna} and initially the flux force is in counter-clock wise direction, as shown in Figure \ref{betaflux}. Therefore, the dominant population paths from {\it Savanna} to {\it Forest} initially move upward under the action of the flux force compared to the paths based purely on the gradient of the landscape. When the grass $G$ decreases close to $\theta_1=0.4$, the savanna sapling-to-adult recruitment rate $\omega(G)$ will run into a threshold value and increase sharply. Therefore, the white path lines arrive at their inflection point, where the flux changes direction to be clockwise. As a result, the white path moves downwards from then on under the action of the flux force, compared to the paths based purely on the gradient of the landscape. Similarly, the red path line starts from {\it Forest} state and initially the flux force is in clock wise direction, as shown in Figure \ref{betaflux}. Therefore, the dominant population paths from {\it Forest} to {\it Savanna} states initially move more upward under the action of the flux force. When the grass $G$ increases close to $\theta_1=0.4$ the savanna sapling-to-adult recruitment rate $\omega(G)$ hits its threshold value and decreases sharply. Therefore, the red path lines arrive at their inflection point, where the flux becomes counter clockwise. The red path moves downwards from then on due to the action of the flux. In ecological terms, this finding is quite intuitive. For a forest to establish, trees must first establish as saplings, whereas they need not regress into saplings before dying, such that the composition of the system is quite different as a {\it Forest} opens up compared to when a {\it Savanna} closes over (see \cite{beckett2021} for empirical evidence supporting this mathematical intuition).

The weight of the population path represents the probability
of each route for state switchings under finite fluctuation. It can be used to quantify the likelihood of different routes for the transition between {\it Savanna} and {\it Forest}. The dominant population path probability can be quantified by the population action $A_{po}(\bf{x})$ as shown in Figure S1 ($\beta$, $\mu$, $\theta_1$) and Figure S22 ($\nu$, $\omega_0$, $\omega_1$). Bigger population action denotes lower dominant population path probability since the dominant population path probability is proportional to $\exp[-A_{po}(\mathbf{x})]$. As $\beta$ becomes larger, the probability of the dominant population path from {\it Forest} to {\it Savanna} decreases, while the probability of the dominant population path from {\it Savanna} to {\it Forest} increases. The variation of the population-potential landscape, the flux, the population actions and the dominant population paths in $\mu, \theta_1, \nu, \omega_0, \omega_1$ are presented in the SI.

\subsubsection*{1.1.b. The intrinsic potential landscape and flux velocity in the zero-fluctuation limit}
\begin{figure}[!ht]
\centering
\includegraphics[width=0.8\textwidth]{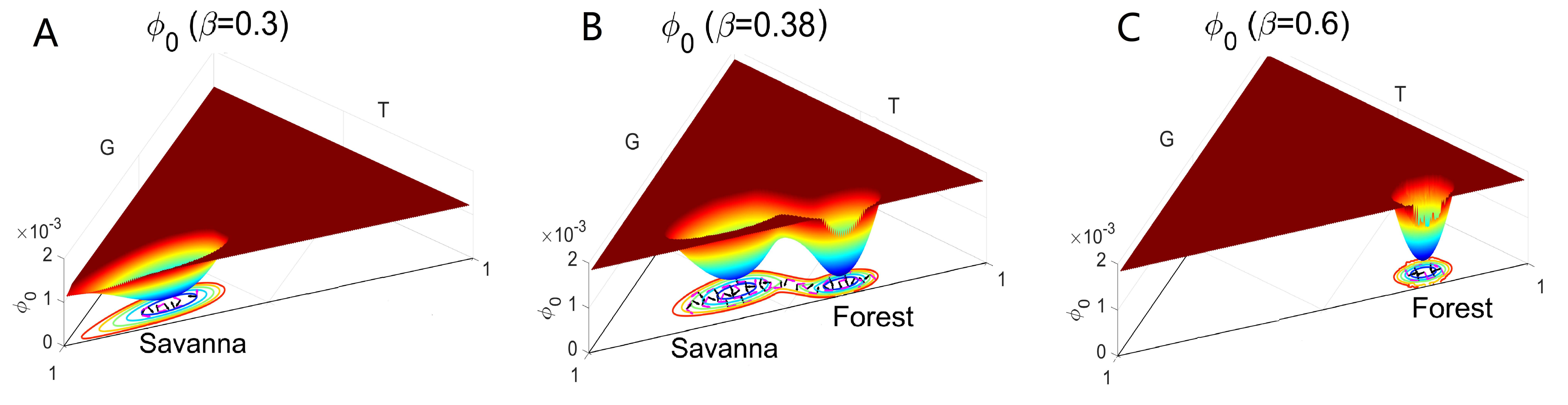}
\caption{The three-dimensional intrinsic potential landscape $\phi_0$ for increasing $\beta$ in zero fluctuation limit. The projection of the flux velocity (purple arrows) and the negative gradient of the intrinsic potential landscape $-\nabla \phi_0$ (black arrows) on the intrinsic potential landscape $\phi_0$ for increasing $\beta$.}\label{phi0_b3d}%
\end{figure}

Figure \ref{phi0_b3d} shows the three-dimensional non-equilibrium intrinsic potential landscape $\phi_0$ with increasing $\beta$. The intrinsic potential landscape changes from a dominant {\it Savanna} stable state to a {\it Savanna} and {\it Forest} coexisting stable state, and then to a dominant {\it Forest} stable state as $\beta$ increases. The intrinsic flux and the negative gradient of the intrinsic potential landscape $- \nabla \phi_0$ are in fact perpendicular to each other and the two dominant intrinsic paths, calculated under the zero fluctuation limit, both pass through the saddle point denoted by the black dot; these facts can also be seen clearly in Figure S2 via a 2D projection.

The dominant intrinsic path probability can be quantified by the intrinsic action $A_{in}(\bf{x})$ shown in Figure S3 ($\beta$, $\mu$, $\theta_1$) and Figure S25 ($\nu$, $\omega_0$, $\omega_1$). The results of the intrinsic actions have the same tendencies with those of the population actions.

Throughout the main text and in the first 6 sections of the SI, the diffusion matrix $\mathbf{G}$ is an isotropic and homogeneous diagonal matrix. In section 7 of the SI, we present results for certain anisotropic and inhomogeneous fluctuations characterized by different choices of diffusion matrices. We perform a coordinate transformation from a special diffusion matrix in an isosceles triangle into an inhomogeneous diagonal matrix in a square (see Figure S40). Thus, we can numerically solve the Hamilton-Jacobi equation in a regular square shape with the resulting diagonal matrix \cite{Zhang2012JCP,Xu2014PLOSONE}. Results for the anisotropic and inhomogeneous fluctuations are qualitatively similar to those with the isotropic and homogeneous diffusion matrices.

\subsubsection*{1.2.Barrier heights and kinetic rates of switching between states}
\begin{figure}[!ht]
\centering
\includegraphics[width=0.8\textwidth]{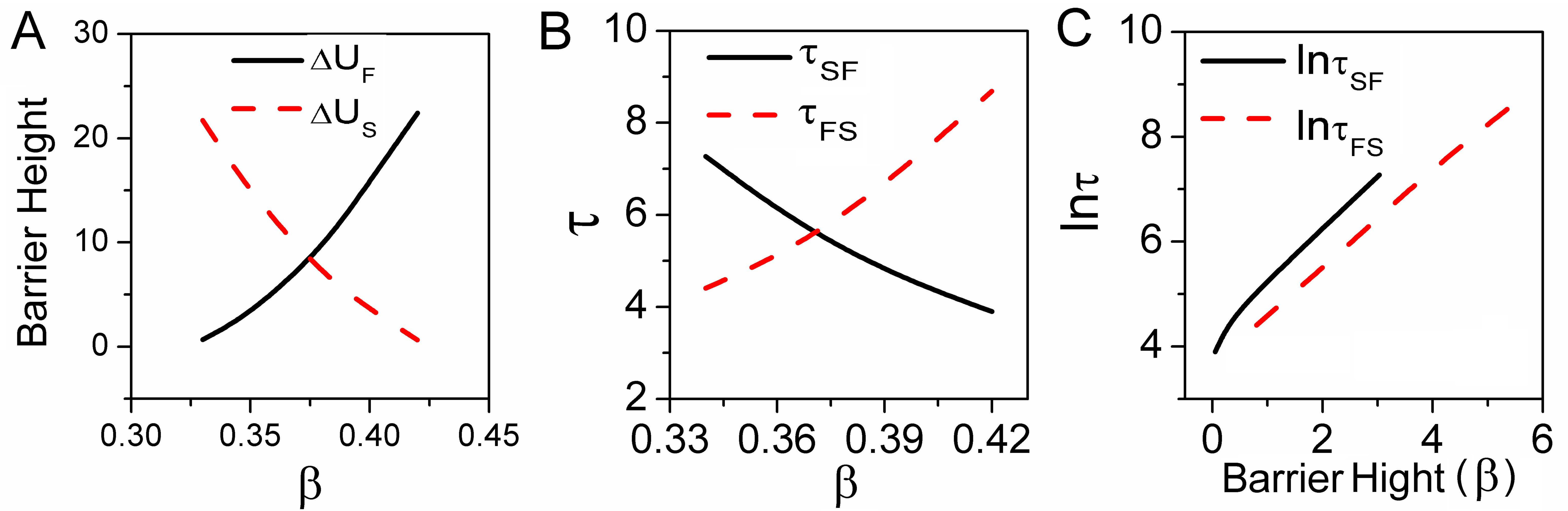}
\caption{A: The population barrier heights versus parameter $\beta$. B: The logarithm of MFPT versus $\beta$. C: The logarithm of MFPT versus barrier heights for $\beta$.}\label{barrier}%
\end{figure}

Figure \ref{barrier}A shows the barrier heights of the population-potential landscape under finite fluctuations as a function of $\beta$. The barrier heights of the intrinsic potential landscape under zero fluctuations
versus $\beta$ are shown in Figure S5A. $\Delta U_F$ represents the barrier height from {\it Forest} to {\it Savanna} and $\Delta U_S$ represents the barrier height from {\it Savanna} to {\it Forest}. $\Delta \phi_{0F}$ represents the intrinsic barrier height from {\it Forest} to {\it Savanna}, while $\Delta \phi_{0S}$ represents the intrinsic barrier height from  {\it Savanna} to {\it Forest}.

As expected, population barrier height $\Delta U_F$ and intrinsic barrier height $\Delta \phi_{0F}$ increase, while $\Delta U_S$ and $\Delta \phi_{0S}$ decrease as $\beta$ increases. High barrier height from the bottom of the basin of the attraction to the barrier top implies that it is difficult to escape from the basin of attraction of that state. Therefore, larger $\Delta U_F$ indicates that the {\it Forest} state is more stable while larger $\Delta U_S$ indicates a more stable {\it Savanna} state. As the sapling birth rate $\beta$ increases, the trees become more established and the {\it Forest} state becomes more stable, while the {\it Savanna} state becomes less stable. The forest-savanna system thus switches from {\it Savanna} dominance to {\it Forest} dominance for $\beta$ sufficiently large. The barrier heights of the population-potential landscape and the intrinsic barrier heights of the intrinsic potential landscape have almost the same qualitative features (cf. Figure \ref{barrier}A and Figure S5A). The barrier heights of the population-potential landscape and the intrinsic potential landscape as functions of $\beta$, $\mu$, $\theta_1$, $\nu$, $\omega_0$, $\omega_1$ are shown in Figures S4 and S5.

Due to stochastic fluctuations or other external forces, ecological systems may not stay in the basin of attraction of their current stable state, but may escape from this basin, switching the system to an alternative stable state. Mean first passage time (MFPT) is the average time for a stochastic process to reach a given threshold value (state) for the first time. The MFPT can be used to quantify the kinetic speed or kinetic time for switching from one state to another, both natural measures for the tendency of the system to escape its current basin of attraction. We use Langevin dynamics to simulate the stochastic SL model and record the MFPT from one stable state to another. We fix a stable state as the starting point, as well as a small area around the other state as the end point, and then record the hitting time when the system enters the small area near the final state. We simulated $40000$ first passage times before averaging to obtain estimates of the MFPT. $\tau_{SF}$ is the MFPT from {\it Savanna} to {\it Forest} while $\tau_{FS}$ is the MFPT from {\it Forest} to the {\it Savanna}. We show the logarithm of MFPT versus $\beta$ in Figure \ref{barrier}B and observe that $\ln\tau_{SF}$ decreases and $\ln\tau_{FS}$ increases as $\beta$ increases. In other words, it becomes easier to switch from {\it Savanna} to {\it Forest} and harder to switch from {\it Forest} to {\it Savanna} as the sapling birth rate increases.

The population-potential landscape topography, quantified by the barrier height, and the corresponding logarithm of MFPT have positive correlation are shown in Figure \ref{barrier}C. Thus, from the barrier height $\Delta U_F$,  $\Delta U_S$ and the logarithm of the corresponding MFPT, $ ln \tau_{FS}$ , one can see that the MFPT is directly related to $\Delta U_F$. Thus, $ ln \tau_{SF}$ has a correlation with the barrier height $\Delta U$ as $\tau \sim exp(\Delta U)$. This shows that
the higher barrier height or the deeper valley is , the longer time it takes to escape from the valley.
This indicates that the population-potential landscape topography
is often correlated to the kinetic speed of the state switching and therefore the communication capability for the global stability of the ecological system. We can also see that kinetic speed can sometimes be estimated based on the population-potential landscape topography through the barrier heights. The logarithm of MFPT versus other parameters and the logarithm of MFPT versus barrier heights are presented in Figure S6 ($\beta$, $\mu$, $\theta_1$) and Figure S26 ($\nu$, $\omega_0$, $\omega_1$). We also found the barrier height $\Delta U_F$, $\Delta U_S$ and the corresponding MFPT $\ln\tau_{FS}$, $\ln\tau_{SF}$ have the correlation of $\tau \sim \exp(\Delta U)$ (see SI).

\subsubsection*{1.3. Early Warning Signals for Bifurcations}
Many complex systems, from ecological systems to financial markets and climates, have tipping points when the systems evolve into a critical dynamic regime. Predicting the system behavior before it reaches a tipping point is extremely difficult, but recent studies in different fields suggest that common early warning signals may be tracked. For a range of different types of systems, these signals can be used to detect whether the critical threshold is approached. When one phase state changes to another, the dispersed fluctuations that occur in the phase transition from an old state to the new state not only increase in magnitude, but also extend in duration. This lengthening of time is called ``critical slowing down'' in statistical physics. The closer the system is to the critical state of sudden change, the more significant this slowing down will be~\cite{Scheffer2012Science,Scheffer2009Nature}. We briefly illustrate how these concepts and techniques can be applied to the SL model in both the finite fluctuation case and in the zero fluctuation limit.

\subsubsection*{1.3.a. Finite Fluctuations}
There are two phase transition points for this set of parameters, as shown in Figure \ref{beta_phase}A. Figure \ref{beta_phase}B shows the population entropy production rate versus $\beta$. Figure \ref{beta_phase}C shows the population average flux versus $\beta$. As $\beta$ increases, population {\it Flux$_{av}$} and population {\it EPR} both increase markedly before then decreasing slightly. Furthermore, both the population averaged flux and population entropy production rate undergo significant changes at the saddle-node bifurcation shown in Figure \ref{beta_phase}A. When undergoing a critical transition at the saddle-node bifurcation, the system has two coexisting stable states in the phase transition region emerging from one stable state. For nonequilibrium systems, while the gradient force always tends to stabilize the point attractor, the flux force (due to its rotational nature) will tend to destabilize the existing state, but stabilize the flow between the states. Therefore, there is a possibility of a new state emerging for the purpose of stabilizing the flow between the existing state and this newly formed state. Thus, the stability of the coexisting state is not determined by the individual state but by the associations between the two states. In this sense, the two states in the transition region have associations with each other, in contrast to the individual state, and more state space can be explored under fluctuations. The system appears to require more average flux to maintain the coexisting states and their associations, in contrast to when there is only one stable state. This also results in greater thermodynamic cost or dissipation. These effects lead to the peaks in the bifurcation region in {\it EPR} and {\it Flux$_{av}$}. We show the population entropy production rate and the population average flux versus other parameters in the SI. Both the population averaged flux and the population entropy production rate have significant changes near (between) the two saddle-node bifurcations under different parameters and their associated changes, reaching similar conclusions as above.

We also found that the {\it Forest} has higher {\it EPR} and {\it Flux$_{av}$}, which implies that the trees need more vegetative growth factors, and more nutrition and energy than grass from the savanna environment. {\it Savanna} needs less vegetative growth factors, less nutrition and less energy than trees from the environment. The population {\it Flux$_{av}$} and population {\it EPR} have similar changes with respect to $\beta$, as shown in Figure \ref{beta_phase}BC. Hence population {\it Flux$_{av}$} and {\it EPR} may provide warning signals of bifurcations. Therefore we may use the population {\it Flux$_{av}$} and {\it EPR} to explore the global stability and bifurcations of the non-equilibrium ecological dynamics.

\begin{figure}[!ht]
\centering
\includegraphics[width=0.8\textwidth]{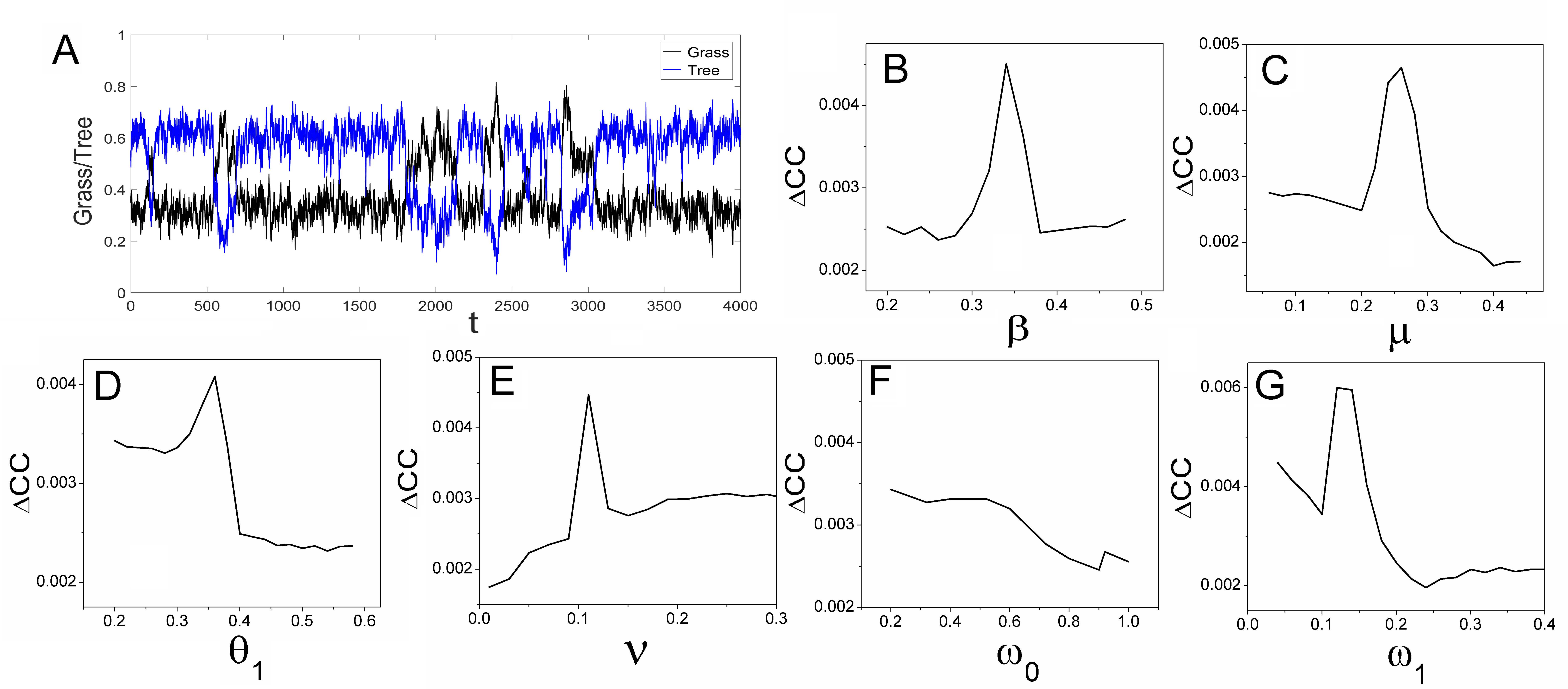}
\caption{A.The trajectories of the {\it Grass} and {\it Tree}. The average change of the forward and backward in time cross correlation function $\Delta CC$ as a function of  different parameters. B: $\beta$, C:$\mu$, D: $\theta_1$, E:$\nu$, F:$\omega_0$, G:$\omega_1$.}\label{corr}
\end{figure}

Long-time trajectories of the SL model with the default values of the parameters and $D=0.0005$ are shown in Figure \ref{corr}A to illustrate this noise induced attractor switching from a pathwise perspective. The average of the differences between the two-point cross correlations forward and backward in time can be used to measure the time irreversibility and therefore the degree of detailed balance breaking ~\cite{Qian2004-2828,ZhangKun2018JPCB,Xu2020JPCB}. The cross correlation function is defined as:
\begin{equation}
C_{XY} (\tau) = \langle X(0)Y(\tau)\rangle =\sum X^iY^jP_i^{ss} P_{ij}(\tau)
\end{equation}
where $X$ and $Y$ denote the time trace signals of variables $X$ and $Y$. $P_i^{ss}$ represents the steady state probability at state $i$ and $P_{ij}(\tau)$ represents the probability from state $i$ to state $j$ with time interval $\tau$~\cite{Qian2004-2828,ZhangKun2018JPCB}. The nonequilibrium steady state probability flux, $J_{ij}^{ss}$, is defined as: $J_{ij}^{ss}=P_i^{ss} k_{ij} - P_j^{ss} k_{ji}$, since $\tau P_{ij} (\tau) \sim k_{ij} \tau$ for small time interval $\tau$, where $k_{ij}$ denotes the transition rate from state $i$ to state $j$, while $k_{ji}$ denotes the transition rate from state $j$ to state $i$. The difference between the forward cross correlation function $C_{XY}(\tau)$ in time and backward cross correlation function $C_{YX} (\tau)$ in time is given by~\cite{Qian2004-2828,ZhangKun2018JPCB}: $C_{XY} (\tau)- C_{YX} (\tau)=X^A Y^B [P_A^{ss} P_{AB}(\tau) - P_B^{ss}P_{BA}(\tau)]=X^A Y^B J_{AB}^{ss} \tau$, where $J_{AB}^{ss}   = \frac{1}{X^A Y^B } \lim_{\tau\rightarrow 0} \frac{C_{XY} (\tau)- C_{YX} (\tau)}{\tau}$.

The difference between the cross correlation functions forward in time and the backward in time can quantify the time irreversibility and flux in ecological systems. We use the average difference in cross correlations $\Delta CC= \sqrt{ \frac{1}{t_f} \int_0^{t_f}  (C_{XY}(\tau)-C_{YX}(\tau))^ 2d\tau }$, which can measure the difference in cross correlation functions between the forward in time and the backward in time with different sets of parameters as shown in Figure \ref{corr}B: $\beta$, C:$\mu$, D: $\theta_1$, E:$\nu$, F:$\omega_0$, G:$\omega_1$. The forward in time cross correlation function and the backward in time cross correlation function are equal to each other with zero flux~\cite{ZhangKun2018JPCB}. Thus, the average difference of cross correlation functions $\Delta CC$ can be used to quantify the flux and therefore the degree of the time irreversibility or the detailed balance breaking~\cite{Qian2004-2828,ZhangKun2018JPCB,Xu2020JPCB}. Importantly, this provides a practical method to quantify one part of the dual driving force of the ecological system from cross correlations between the real time trajectories of the empirical observations of the variables of the interests in ecological systems. Along with the earlier finding in this study, we can quantify both driving forces of the ecological system: the landscape from the frequency statistics or distribution of the variables and the flux from cross correlations between real time trajectories from the empirical observations of ecological dynamics. As detailed in the SI, the average difference of cross correlations has almost the same trends with respect to different parameters as the {\it Flux$_{av}$} and {\it EPR}. Crucially, the average differences in cross correlations become significantly higher near (between) the two saddle-node bifurcation region. This provides a possible practical way to infer the onset or offset of the bifurcation from the observed time traces of the ecological systems, giving rise to a possible early warning signal.

\begin{figure}[!ht]
\centering
\includegraphics[width=0.8\textwidth]{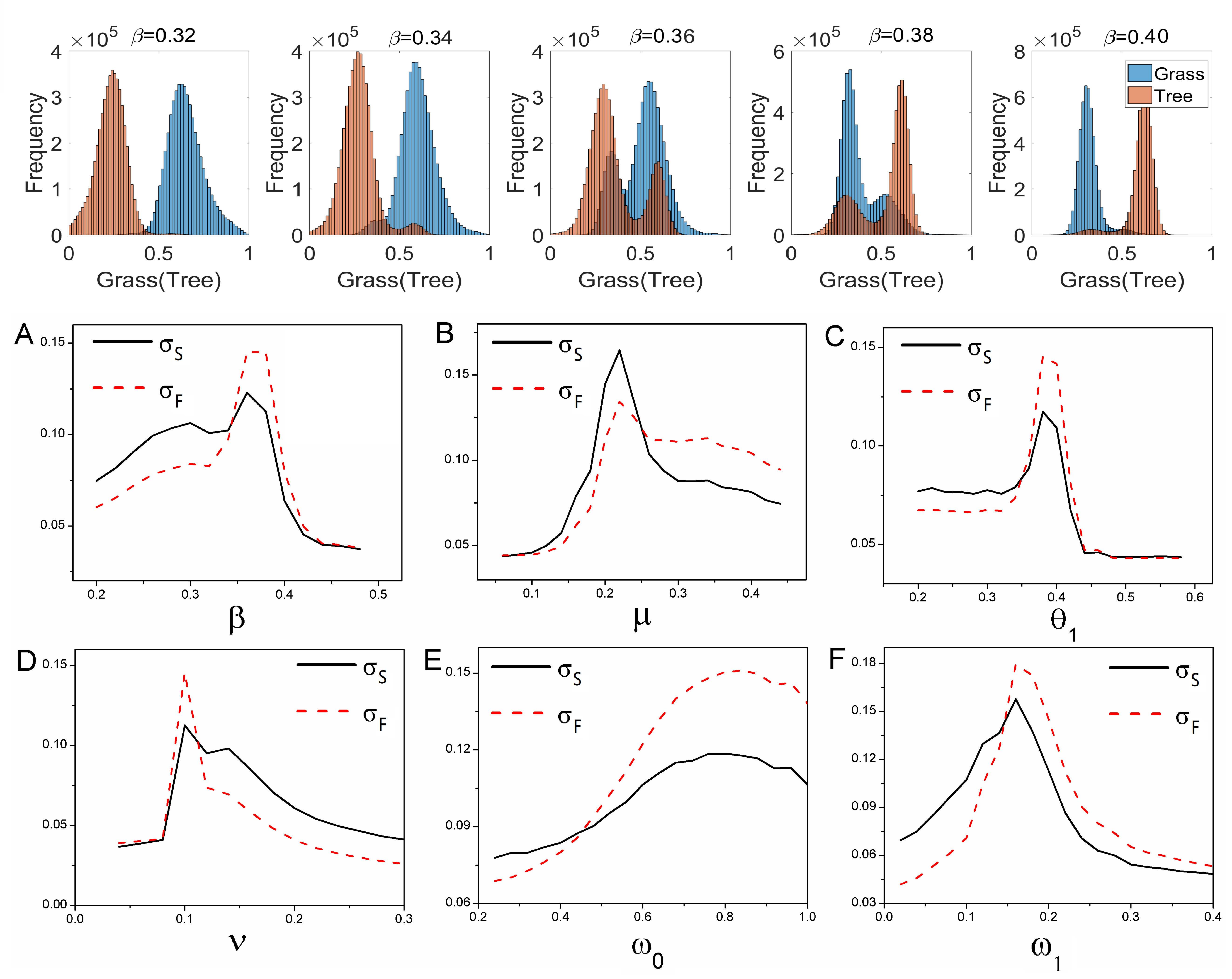}
\caption{The frequencies of {\it Grass} and {\it Tree} along the variable state space with $\beta=0.32,0.34,0.36,0.38,0.40$ in the above panel. The variances {\it Grass} $\sigma_S$ and the variances {\it Tree} $\sigma_F$ versus A: $\beta$, B:$\mu$, C: $\theta_1$, D:$\nu$, E:$\omega_0$, F:$\omega_1$.}\label{Pdis_sig}
\end{figure}

The distributions of {\it Grass} and {\it Tree} in the state space are shown for various values of $\beta$ in Figure \ref{Pdis_sig} and we observe the expected change from {\it Savanna} to {\it Forest} dominance as $\beta$ increases. The variances of {\it Grass} and {\it Tree} versus various parameters are also shown in Figure \ref{Pdis_sig}. The variances of {\it Grass}, $\sigma_S$, and the variances {\it Tree}, $\sigma_F$, all have peaks near (between) the two saddle-node bifurcations for each parameter. Thus the variances from
the empirical observations of the ecological dynamics can also
provide a possible early warning signal for the onset of a bifurcation.

\begin{figure}[!ht]
\centering
\includegraphics[width=0.8\textwidth]{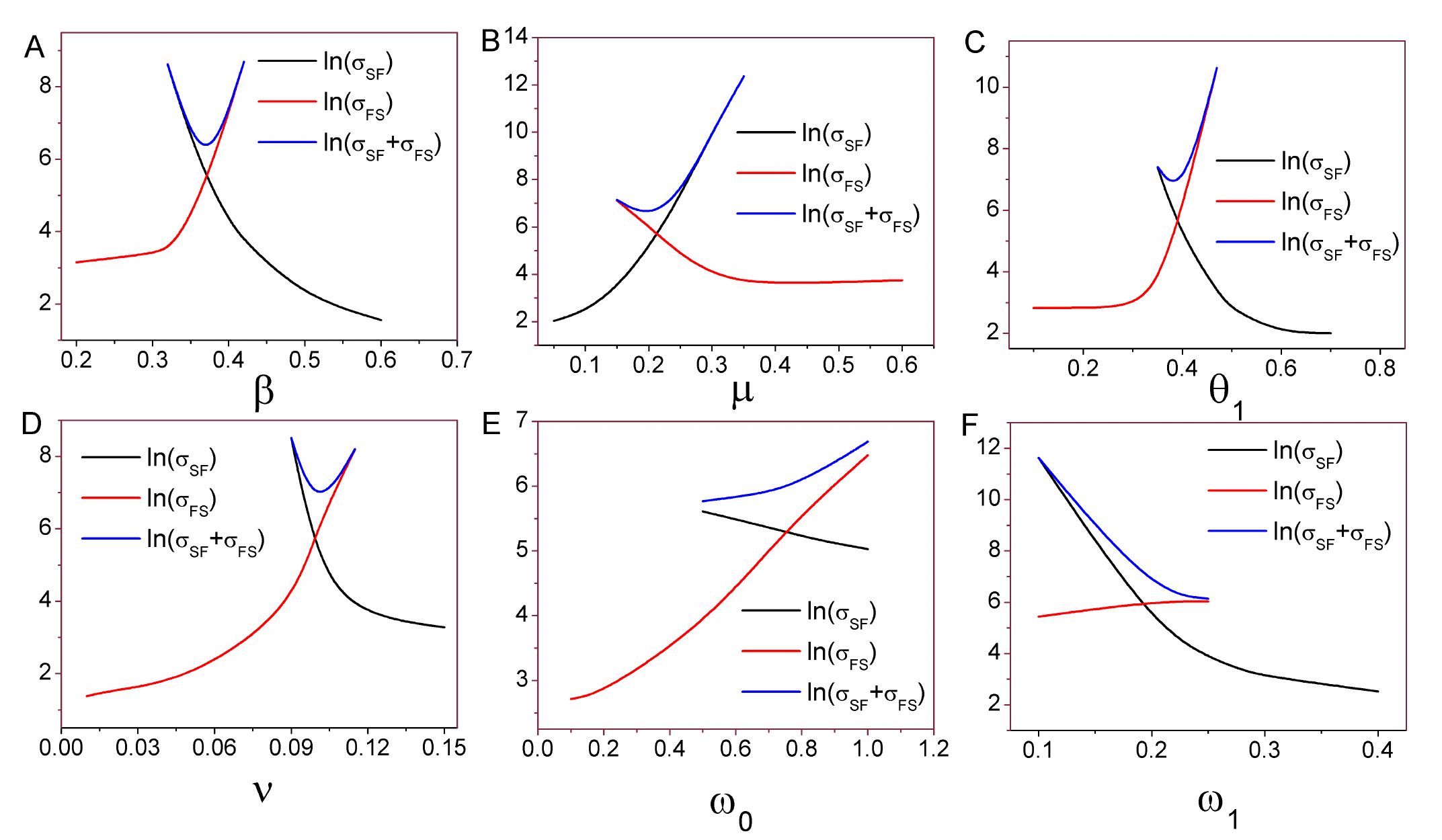}
\caption{The logarithms of the variances of the first passage time from {\it Savanna} to {\it Forest} $\log(\sigma_{SF})$ and the first passage time from {\it Forest} to {\it Savanna} $\log(\sigma_{FS})$, and the logarithms of the sum of them $\log(\sigma_{SF}+\sigma_{FS})$. A: $\beta$, B: $\mu$, C: $\theta_1$, D: $\nu$, E:  $\omega_0$, F: $\omega_1$.}\label{FPT_sig}
\end{figure}

The logarithms of the variances of the kinetic first passage time from {\it Savanna} to {\it Forest} $\log(\sigma_{SF})$ and the logarithms of the variances of the kinetic first passage time from {\it Forest} to {\it Savanna} $\log(\sigma_{FS})$, and their sum $\log(\sigma_{SF}+\sigma_{FS})$ are shown in Figure \ref{FPT_sig}. For one stable {\it Savanna} state, we calculate the first passage time from a certain point near the {\it Forest} state determined in the two state parameter regime to the {\it Savanna} state. In reverse, we can obtain the first passage time for one stable {\it Forest} state. We can see that both $\log(\sigma_{SF})$ and $\log(\sigma_{FS})$ monotonically increase or decrease as the parameter changes. This indicates that the more stable the states are, the larger the variances of the first passage time will be. The sums of the variances have a ''U'' shape, indicating that when the ecological system has two stable state coexisting, there are fewer fluctuations in its kinetics. This is because the communication between the two states in the two state coexisting case is much more frequent than the one stable state alone. While the gradient always tends to stabilize the point attractors, the flux tries to destabilize the point attractor, but stabilizes the flow between the stable states. Thus, the flux flow between the two stable states drives the stability of the state coexistence. The sum of the variance of the kinetic first passage times forward and backward has significant changes near (between) the two saddle-node bifurcation regime. This shows that through the real time trace analysis of the ecological dynamics, one can use the fluctuations in kinetics to quantitatively locate where the bifurcation will be likely to occur, thus providing a possible early warning signal.

\subsubsection*{1.3.b. The Zero Fluctuation Limit}
\begin{figure}[!ht]
\centering
\includegraphics[width=0.8\textwidth]{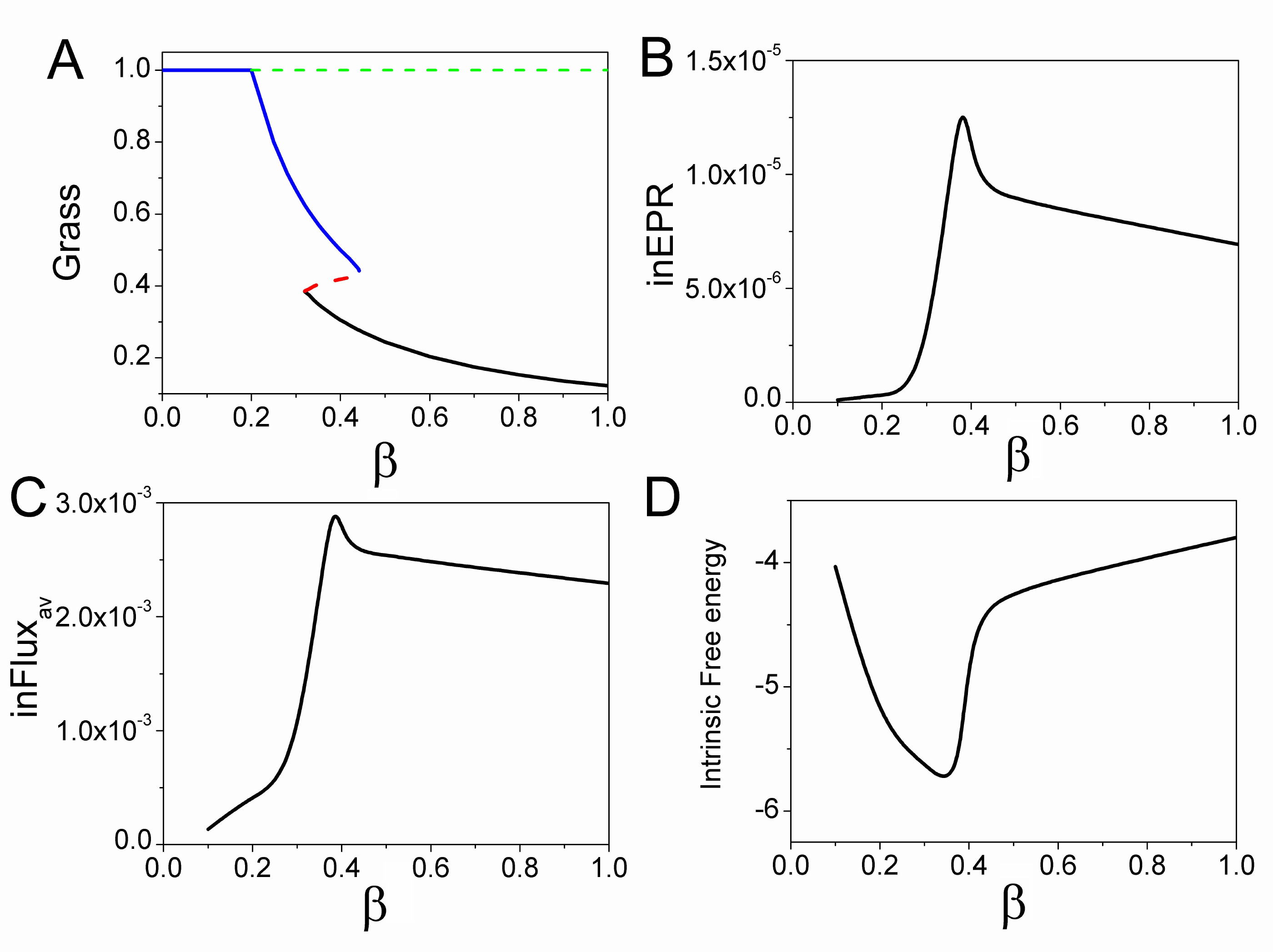}
\caption{A: The phase diagram  versus $\beta$. B: The intrinsic entropy production rate versus $\beta$. C: The intrinsic average flux versus $\beta$. D: Intrinsic free energy versus $\beta$.
}\label{phi0_betahe}%
\end{figure}

There are two phase transition points for the set of parameters we considered, as shown in Figure \ref{phi0_betahe}A. Figure \ref{phi0_betahe}A is the same as Figure \ref{beta_phase}A since it is convenient to check the transition zone and compare with the other subfigures. Figure \ref{phi0_betahe}B shows the intrinsic entropy production rate versus $\beta$, while Figure \ref{phi0_betahe}C shows the intrinsic average flux versus $\beta$. We can see that as $\beta$ increases, intrinsic {\it EPR} and intrinsic {\it Flux$_{av}$} both increase first, then decreases. We can see both the intrinsic averaged flux and intrinsic entropy production rate have significant changes along with the bifurcation shown in Figure \ref{phi0_betahe}A. We also found that the {\it Forest} has more intrinsic {\it EPR} and intrinsic {\it Flux$_{av}$}, which implies that the trees need more vegetative growth factors, and cost more nutrition and energy from the savanna environment. While {\it Savanna} needs less vegetative growth factors, and costs less nutrition and energy from the savanna environment. We can see that the peaks in the intrinsic {\it EPR} and the intrinsic {\it Flux$_{av}$} are distinct at the phase transition region. This shows that the intrinsic {\it EPR} and the intrinsic {\it Flux$_{av}$} may be used to characterize the fundamental properties of the system, such as bifurcations. Both the intrinsic averaged flux and the intrinsic entropy production rate under zero fluctuations have more significant changes near (between) the two saddle-node bifurcation regions than those of population entropy production and population flux under finite fluctuations.

However, the critical slowing down theory only applies to the continuous phase transition (often called second order phase transition) and hence for the discrete phase transition (such as first order phase transition), we can not use critical slowing down to trace the bifurcation signals. From this study and some others~\cite{Xu2020JPCB}, {\it EPR} and {\it Flux$_{av}$} can both be used as possible early warning signals or markers for bifurcations such as subcritical pitchfork bifurcation and supercritical-pitchfork bifurcation, as well as saddle-node bifurcation. Both {\it EPR} and {\it Flux$_{av}$} are not constrained to only be markers for continuous transitions. In fact, they apply to both discrete and continuous transitions in the examples shown. We suggest that the above statement may be general. Therefore, this may provide a feasible way to predict the bifurcations in ecological systems to avoid catastrophic change.

From the above discussions, {\it EPR} and {\it Flux$_{av}$} are observed to have extreme values between the two saddle-node bifurcations as the parameter varies.  The locations of these peak values in {\it EPR} and {\it Flux$_{av}$} are close to the place where the two stable states are equal in chances (probabilities) of appearance (Figure \ref{betaU2d}, Figure \ref{phi0_b3d}, and Figure \ref{phi0_betahe}). This is because the emergence of two stable states and their associate connections needs more consumptions to maintain in contrast to that of the one individual state.  Away from the extreme values in average flux and entropy production or equal probability of the two states, the system switches the dominance of the stability on the landscapes (Figure \ref{betaU2d} and Figure \ref{phi0_b3d}). Thus, near the regime of the coexistence of the two stable states, the former stable state becomes less stable while the former less stable state becomes more stable when the system goes through the regime with extreme values of {\it EPR} and {\it Flux$_{av}$} or equal basin depths of the two stable states Figure \ref{betaU2d} and Figure \ref{phi0_b3d}). It will then be easier to switch  from the former stable state to the latter stable state. Since the {\it EPR} and {\it Flux$_{av}$} change significantly upwards before their peak values, they provide possible early warning signals for the bifurcations or more explicitly the switching in the dominance of the stabilities from one state to another (Figure \ref{betaU2d}, Figure \ref{phi0_b3d}, and Figure \ref{phi0_betahe}).

Figure \ref{phi0_betahe}D shows the intrinsic free energy versus $\beta$. The slope of the non-equilibrium intrinsic free energy significantly changes near the phase transition zone, although the non-equilibrium intrinsic free energy is continuous. Figure \ref{phi0_betahe}B, Figure \ref{phi0_betahe}C and Figure \ref{phi0_betahe}D show that significant slope changes in intrinsic entropy production rate, intrinsic average flux and intrinsic free energy (analogous to the equilibrium case) may provide signals of bifurcation. We can see that the non-equilibrium intrinsic free energy may also be useful to quantify the global phases of the system and the bifurcations. Therefore, we may use this non-equilibrium intrinsic free energy function as well as the intrinsic {\it Flux$_{av}$} and intrinsic {\it EPR} to explore the global stability and bifurcations of the non-equilibrium ecological dynamics.

\subsection*{2.Stochastic Fluctuations generate a new Stable State}
\begin{figure}[!ht]
\centering
\includegraphics[width=0.80\textwidth]{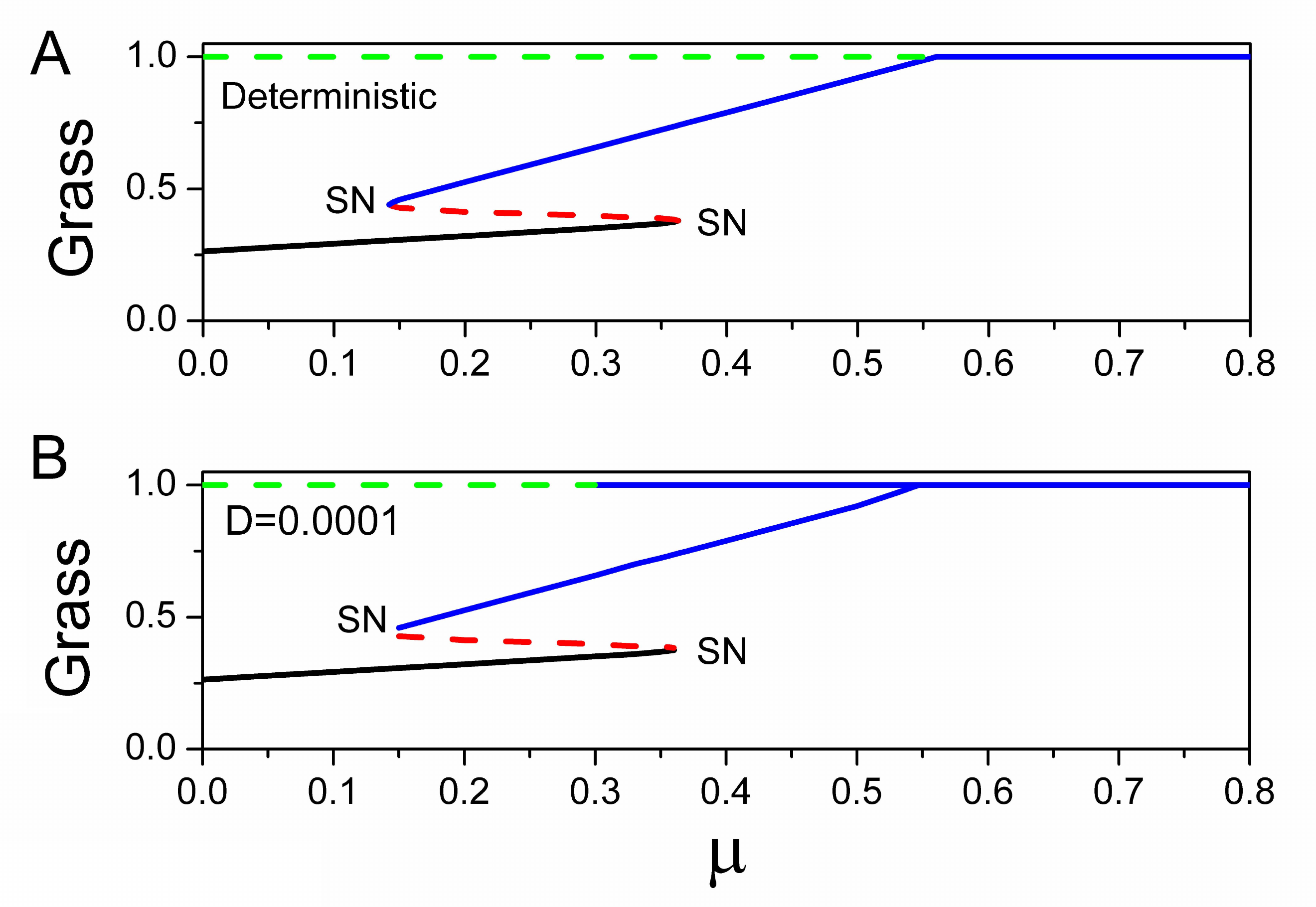}
\caption{A: Deterministic phase diagram in $\mu$. B: Stochastic phase diagram (minima of the population-potential landscape) in $\mu$ with $D=0.0001$.}\label{miuDphase}%
\end{figure}

Figure \ref{miuDphase}A show the deterministic phase diagram mapping the fraction of Grass cover versus $\mu$. When the savanna saplings mortality rate is very small, the only stable state is {\it Forest}, but as $\mu$ is increased, the dynamics shifts from {\it Forest} to {\it Savanna}. For $\mu$ between about $0.14$ and $0.37$, bistability emerges and beyond $0.37$, the dynamics approach a state dominated by Grass. As $\mu$ increases further, trees occasionally go extinct and a new quasi-stable {\it Grassland} state emerges. For $\mu$ larger than $0.37$ but less than $0.65$, the {\it Grassland} state is always stable on the G-axis and unstable on the T-axis. For $\mu$ beyond $0.65$, this state becomes stable. Figure \ref{miuDphase}B shows the corresponding stochastic phase diagram versus $\mu$ with $D=0.0001$. Remarkably, the {\it Grassland} state becomes quasi-stable much sooner than stable in the deterministic phase diagram (around $\mu = 0.3$) in the presence of stochastic fluctuations.

\begin{figure}[!ht]
\centering
\includegraphics[width=0.8\textwidth]{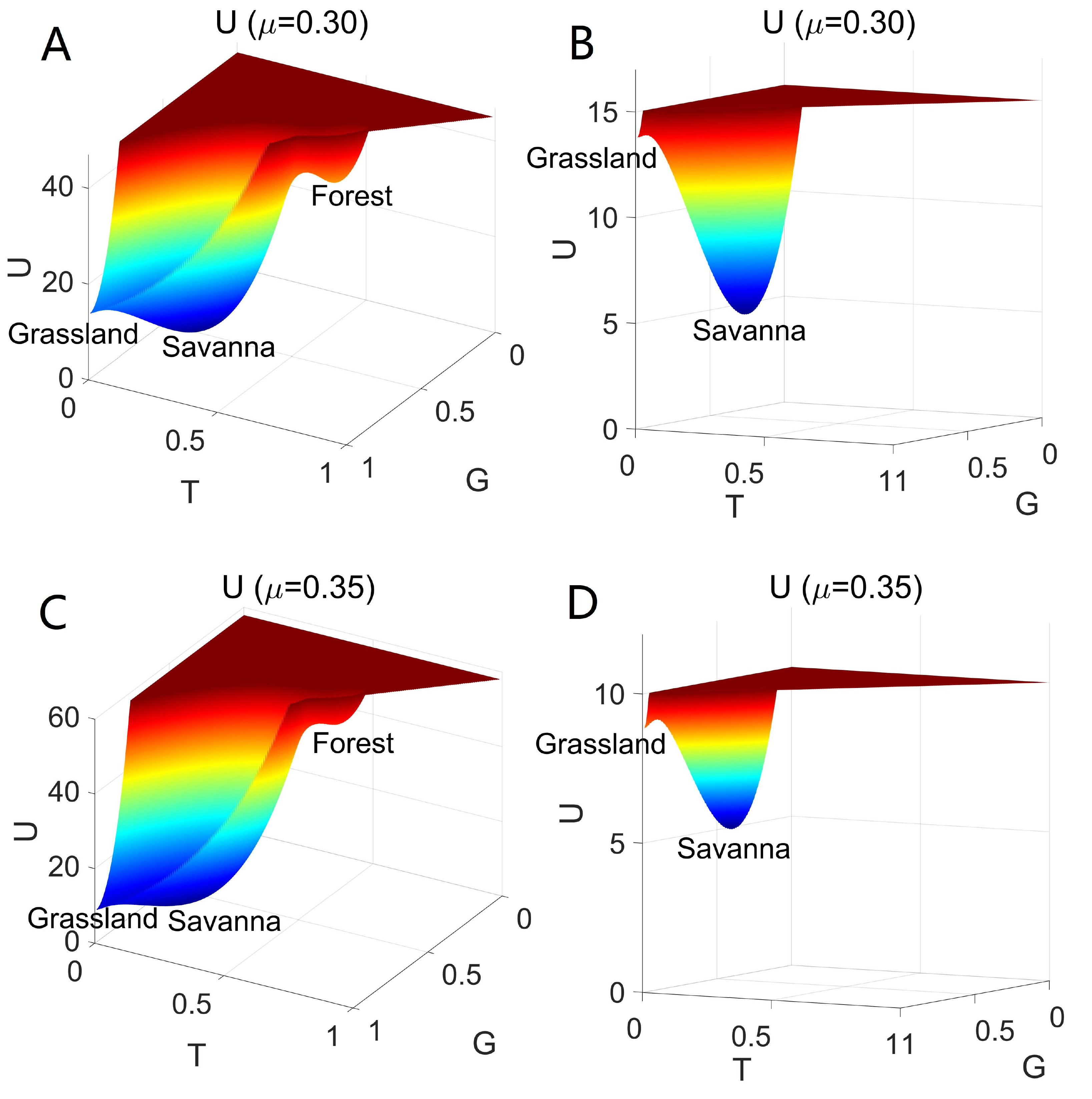}
\caption{ The three-dimensional population-potential landscapes versus A: $\mu=0.3$, C: $\mu=0.35$. The figures on right panel are the magnification basin of the left panel with lower maximum values.}\label{mu3dlarge}%
\end{figure}

Figure \ref{mu3dlarge}AC shows three-dimensional population-potential landscapes varying with increasing savanna saplings mortality rate $\mu$ on the left panel with $D=0.0001$. The right-hand panels (Figure \ref{mu3dlarge}BD) are the magnification of the left-hand panel with lower cutting maximum values. The right subfigures show the details around the {\it Grassland} state $[1,0]$. For $\mu=0.3$, as in Figure \ref{mu3dlarge}A, the system has two stable fixed points in its deterministic phase diagram (Figure \ref{miuDphase}A) and three stable fixed points in its stochastic phase diagram (shown in Figure \ref{miuDphase}B); the {\it Grassland} state $[1,0]$ is not a stable fixed point for the deterministic SL model for this value of $\mu$. We can see from Figure \ref{mu3dlarge}D that the {\it Grassland} state is a stable state with a very small basin in stochastic dynamics when $\mu=0.3$. Figures \ref{mu3dlarge}C and \ref{mu3dlarge}D show the population-potential landscape with $\mu=0.35$. The population-potential landscape around $[1,0]$ shows that {\it Grassland} state has a relatively small basin of attraction, but this basin is larger than that shown in Figure \ref{mu3dlarge}A, indicating that the basin of the {\it Grassland} state becomes deeper as $\mu$ increases.

We emphasize that the emergence of the {\it Grassland} state in Figure \ref{mu3dlarge}, leading to the coexistence of the three stable states {\it Forest}, {\it Savanna} and {\it Grassland}, contrasts with the deterministic dynamics that predicts the absence of forest extinction. The newly quasi-stable {\it Grassland} state is thus born from the stochastic fluctuations. An important prediction of the landscape-flux approach is the emergence of new quasi-stable states due to fluctuations. Figure \ref{mu3dlarge} highlights this effect, where the deterministic system predicts the existence of only two stable states ({\it Forest} and {\it Savanna}, see Figure \ref{miuDphase}A) and the stochastic system displays an additional quasi-stable {\it Grassland} state (Figure \ref{miuDphase}B and Figure \ref{mu3dlarge}, for $0.30< \mu<0.37$). {\it Grassland} is thus an unstable fixed point in deterministic dynamics which is stabilized by the fluctuations, and a basin of attraction emerges in its vicinity (see magnification around the of left panel). As $\mu$ increases, the {\it Grassland} basin becomes much deeper and thus the associated state becomes more attractive while the {\it Savanna} state becomes a much shallower minimum of the population-potential landscape, and thus less stable.

In the deterministic case, the {\it Grassland} state $[1,0]$ is stable in the $G$ direction and unstable in the $T$ direction. However, the constraint $G+T+S=1$ and the boundary conditions $\{G>0, T>0\}$ effectively mean there are three reflecting walls so that $G>0, T>0$ and $G+T+S=1$ constrain the system to this specific region. In particular, at $[1,0]$, the deterministic system is stable in the $G$ direction but also not freely movable in the $T$ direction. The noise effectively expands the accessible region near the point $[1,0]$. Since there are two walls with infinite potential at $T=0$ and $G+T =1$ near $[1,0]$, the population-potential landscape in the T direction must go up towards both the $T=0$ wall and the $G+T=1$ wall. This leads to an effective quasi-stable region or basin of attraction in the $T$ direction, in addition to that already present in the G direction near $[1,0]$. Thus, the phase space around $[1,0]$ changes from a saddle point region to a stable basin due to the dual actions of the noise and boundary conditions.
\subsection*{Conclusion and Discussion}
The dynamics of ecological systems are determined by both the population potential, which drives the system towards the potential minimum, and the curl flux, which describes the switching dynamics between basins of attraction. The population-potential landscape can be obtained from the frequency distribution of the observables at long times and can provide insights on the global stability. The steady-state probability flux has divergent free curl nature and provides the driving force for the non-equilibrium part of the dynamics. Stability is crucial for exploring the function and robustness of an ecological system, but this is a challenging issue both theoretically and practically. Ecological stability is commonly defined as Lyapunov stability, which can describe the global stable behavior of the ecological system under perturbations. However, it is difficult to obtain a Lyapunov function for complex ecological systems. Here, a general method is introduced to calculate the underlying intrinsic potential landscape as a Lyapunov function that can quantify the global stability of the ecological system. In particular, the intrinsic potential landscape $\phi_0$ is shown to be a Lyapunov function in the zero noise limit. In fact, $\phi_0$ monotonically decreases along the deterministic trajectory towards the global minima and therefore this leads to an optimization principle for ecological systems which can characterize the global stability.

We investigate the SL model with the underlying potential-flux landscape. We found many interesting quantitative markers for the stability and dynamics: barrier heights between the basins of attractions, mean first passage time MFPT representing the kinetic time of state switching, the entropy production rate {\it EPR} representing the thermodynamic cost, and the average flux representing a dynamical driving force for quantifying the global stability of the forest-savanna ecology system. We found that the average flux {\it Flux$_{av}$} and the entropy production rate {\it EPR} have the same trends. Their significant changes near (between) the two saddle-node bifurcation points. It is particularly evident for zero fluctuation situation with the intrinsic potential landscapes. This demonstrates that the flux and the entropy production rate may provide dynamical origin and thermodynamic origin for bifurcation of the nonequilibrium ecological systems respectively. The results show that forest-savanna systems can be stable and robust under certain environments. The dominant paths describing how the state switching processes actually occur do not follow the naively expected steepest descent gradient path based on the population-potential landscape alone because of the presence of the non-zero flux. In fact, they are irreversible and determined by both the population-potential landscape and the flux. The flux is the source of the irreversibility.

Remarkably, even though the {\it Grassland} state is unstable in the deterministic case without fluctuation for certain parameter ranges, it can in fact become quasi-stable under small fluctuations. Moreover, the emergence of this new stable state under stochastic forcing is predicted by our landscape-flux theory. This study can shed light on how to use the observed frequency distribution to quantify the landscape as one of the two driving forces of ecological dynamics and gain insights on the stability of the ecological system.

We provide quantitative and physical markers for identifying the start and end of a bifurcation via the {\it EPR}, the {\it Flux$_{av}$} and the intrinsic free energy. More practically, the information on the physical bifurcation markers, such as the flux and thermodynamics cost {\it EPR}, can be inferred from the time irreversibility of the observed time traces. Similarly, the variance in the frequency statistics and kinetic time obtained directly from the observed time traces can be used as the kinematic markers for the onset and offset of bifurcations. Therefore, we may be able to identify both physical and kinematic markers to detect the beginning and end of bifurcations in ecological systems based on the observed time series data.

\subsection*{Acknowledgment}
LX thanks supports from NSFC No.11305176, 21721003. JW thanks support
from NSF DMS-1951385; SL and DP thank support from NSF DMS-1951358; CS thanks support from NSF DMS-1951394.

\end{document}